%% file: IEEE_TCOM.tex
\documentclass[journal,draftcls,onecolumn,12pt,twoside]{IEEEtranTCOM}
\usepackage{cite}
\usepackage{amsmath,amssymb,amsfonts}
\usepackage{algorithmic}
\usepackage{graphicx}
\usepackage{textcomp}
\usepackage{xcolor}
\usepackage{nccmath}
\usepackage{tikz}
\normalsize

\usepackage{cite}

\ifCLASSINFOpdf
\else
\fi

\usepackage{array}

\usepackage{mdwmath}
\usepackage{mdwtab}

\usepackage{eqparbox}
\begin{document}
%
\title{An Efficient QAM Detector via Nonlinear Post-distortion based on FDE Bank under PA Impairments}
%
%
%

\author{Murat~Babek~Salman,~\IEEEmembership{Student~Member,~IEEE,}
        and~Gokhan~Muzaffer~Guvensen,~\IEEEmembership{Member,~IEEE}
\thanks{M. B. Salman and G. M. Guvensen are with the Department
of Electrical and Electronics Engineering, Middle East Technical University, Anakara, 06800 Turkey. E-mails: mbsalman@metu.edu.tr, guvensen@metu.edu.tr.}
}

%
%

\markboth{IEEE Transactions on Communications}%
{Submitted paper}
%



\maketitle

\begin{abstract}
In this paper, we propose a novel receiver structure for single-carrier transmission with frequency domain equalization (FDE) that is exposed to power amplifier (PA) nonlinearities. A two-stage approach is adopted, in which linear communication channel is equalized at the first stage, and it is followed by a post-distortion where nonlinear distortion is reduced. In literature, nonlinear processing techniques are proposed, which performs memoryless compensation of nonlinear distortion together with FDE. However, in this study, we show that even if a memoryless nonlinearity exists, the received signal is impaired by nonlinear inter-symbol-interference. Therefore, we propose a class of symbol rate post-distortion techniques, which use neighboring received symbols to suppress the nonlinear interference. Two different post-distortion techniques, Gaussian process regression (GPR) and neural network (NN) based post-distorters, are considered. Also, a decision metric, combining outputs of fractional delayed bank of FDE's after post-distortion, is proposed to overcome performance degradation of FDE for frequency selective channels under nonlinear distortion. Performances of the proposed techniques are compared with that of the state-of-the-art techniques in terms of bit error rate and achievable information rate metrics via simulations. Simulation results demonstrate that GPR and NN based post-distortion methods together with bank of FDE outperform state-of-the-art techniques.

\end{abstract}

\begin{IEEEkeywords}
Nonlinear power amplifiers, nonlinear post-distortion, nonlinear ISI, machine learning based post-distortion, FDE for nonlinear channels.
\end{IEEEkeywords}

%
\IEEEpeerreviewmaketitle

\section{Introduction}

\IEEEPARstart{G}{oal} to achieve high data rates with limited spectral resources makes higher order quadrature amplitude modulation (QAM) constellations attractive for future generation wireless networks \cite{highConst1,highConst2,highConst3}. Unfortunately, performance of such higher order modulations is significantly affected by the non-ideal hardware. For instance, PA's are operated on low backoff regions in order to increase power efficiency and meet transmit power requirements, however, this comes with a price of notable nonlinear distortion. Due to high peak-to-average power ratio (PAPR) of massive constellations, such as $1024$ QAM, signals are subjected to severe nonlinear distortion, which degrades the system performance \cite{highConst3}.

In addition to nonlinearity, intersymbol interference (ISI) due to dispersive wireless channel is an important issue that should be taken into consideration \cite{isi_ch}. In recent communication systems, this problem is solved by using orthogonal frequency division multiplexing (OFDM), which provides reduced complexity frquency domain channel equalization (FDE) compared to single carrier (SC) modulation with time domain equalization. However, due to multiplexing of symbols by inverse discrete Fourier transform (IDFT) operation, time domain signal in OFDM have Gaussian distribution, which yields high PAPR. Thus, OFDM is inherently vulnerable to PA nonlinearities, which makes SC more appealing for systems with nonlinear PA. Application of FDE in SC transmission is studied for single-input single-output (SISO) systems in \cite{sc_fde1,sc_fde2} to reduce the implementation complexity of SC receiver. In addition, SC transmitter does not employ complex IDFT operation; therefore, it is more suitable for uplink channels since computational capabilities are limited at the user terminal (UT). It is known that performance of SC-FDE  degrades significantly due to noise enhancement especially for the frequency-selective channels having deep fades in the spectrum \cite{noise_enhance}. Existence of nonlinear PA at the transmitter degrades the FDE performance even further due to additional nonlinear distortion amplification. However, to the author's knowledge, effects of nonlinear PA on the FDE performance is not studied in literature.

\subsection{Related Literature}
In literature, there are many methods that are proposed for linearization of PA's at the transmitter. Digital predistortion (DPD), which aims to linearize the transmitted signal subjected to nonlinear distortion, is the most popular technique among others \cite{GMP_jour,dpd2,arvtdnn}. DPD is a suitable solution for downlink transmission since base stations (BS) have quite sufficient computational power to implement complex DPD algorithms. However, implementing DPD at UT's is quite costly; thus, techniques avoiding utilization of DPD are more suitable for uplink transmission \cite{uplink_sc1,uplink_sc}. Post-distortion techniques, which operate on nonlinearly distorted received signals, are developed in order to implement complex compensation algorithms at BS \cite{hammer_twc2017,post2,post1,Colavolpe,NN_eq1}. In \cite{post2} and \cite{Colavolpe}, Volterra series based post-distorters are employed as a nonlinear equalizer to detect the transmitted symbols. Besides, in \cite{post1}, Gaussian process regression (GPR) is also adopted as nonlinear equalizer for nonlinear channels to improve the detection performance. However, nonlinear equalizers proposed in \cite{post1,post2,Colavolpe,NN_eq1}, treat nonlinearity together with wireless channel, which increases computational complexity. In addition, decision region optimization based detectors \cite{ziya,SalehDet} are recently proposed to decode nonlinearly distorted symbols. These techniques exploit constellation point dependent distortion, where nonlinear distortion becomes more distinctive as symbol power increases.

Recently, methods that decouple linear and nonlinear channels are proposed in order to reduce computational complexity \cite{hammer_twc2017,hammer_tsignal_2014}. These studies consider the system as a memoryless nonlinearity followed by linear wireless channel. By exploiting Hammerstein channel assumption, first linear wireless channel is equalized by using hybrid decision feedback FDE (HDFDE) then equalized signal is processed by the memoryless post-distorter, which is the estimated inverse of the memoryless nonlinarity. Complex valued B-spline neural network is utilized in order to extract inverse nonlinear model \cite{bspline1,hammer_twc2017,hammer_tsignal_2014}. However, pulse shaping, which creates inherent nonlinear memory, is not considered in these studies. Significance of the memory due to pulse shaping is shown numerically in \cite{ziya,SalehDet} for memoryless PA's. These studies demonstrate that nonlinear ISI yields I/Q correlated distortion, and developed decision metrics that outperform conventional detector by taking the distortion term into account. However, these detectors do not consider memory in detection, they only consider nonlinear ISI as distortion source. In addition, effects of nonlinear distortion on the FDE performance are not considered in these studies since the signal model is formed in symbol-sampled domain, where perfect decoupling of linear and nonlinear systems is possible.

In addition to state-of-the art decision metric, \cite{ziya} and \cite{SalehDet} develop a framework to evaluate performance of the system impaired with transceiver non-idealities. Capacity expression based on generalized mutual information (GMI) metric \cite{Capacity:1} is adopted to obtain mismacted decoding capacity \cite{GMI:1,GMI:2} based on assumed probability density function (PDF) of the received signal. In this study, we also employ mismatched decoding capacity in order to find a lower bound on achievable information rate.

\subsection{Contributions}

In this paper, a novel receiver design is proposed and extensive analysis is carried out for the received signals impaired by transceiver nonlinearities. Firstly, a nonlinear distortion analysis is conducted to investigate the effects of frequency selective channels on the received nonlinear distortion signal. It is concluded that equivalent channel experienced by the distortion signal is different then that of experienced by the desired signal. Consequently, distortion signal is highly amplified when the desired signal experiences deep fade although equivalent channel for the distortion is rather flat.

Secondly, two nonlinear post-distortion methods are proposed as an alternative to sequence detector since sequence detection is NP-hard for nonlinear systems with memory, where PDF of the received signal is not known. Thus, machine learning type algorithms are employed to perform multi-dimensional detection. For this purpose, neural network (NN) and Gaussian process regression (GPR) based post-distortion algorithms are proposed. In literature, GPR based equalizers are proposed for joint equalization of both fast varying linear channel and slowly varying nonlinear channel simultaneously \cite{post1,post2}. However, hyper-parameter optimization for GPR is a time consuming process; therefore, it is more practical to employ GPR to compensate the slowly varying nonlinear effects rather than complete nonlinear channel. Besides, the problems, considered in \cite{post1} and \cite{post2}, are defined in symbol rate for lower order constellations; hence, effects of pulse shaping and high PAPR are not considered. Also note that longer training sequence is needed for massive constellations such as $1024$ QAM. Therefore, decoupling linear equalization and nonlinear post-distortion is crucial to reduce computational complexity. In addition, best linear unbiased estimator (BLUE) \cite{gokhan_hoca_blue} is adopted to further decrease the complexity of GPR parameter optimization.

Lastly, it is observed that performance of the receiver, operating at symbol rate, is significantly affected by the sampling instant when the transmitted signal is nonlinearly distorted especially for dispersive channels. In FDE procedure, vast amount of gain is applied to equalize the frequency bins experiencing deep fades. Consequently, methods employing FDE on nonlinearly distorted signals prior to nonlinear equalization are subject to significant amount of distortion amplification. As a result, block errors occur due to amplified distortion term that cannot be handled by nonlinear post-distortion. On the other hand, experienced effective channel at symbol rate depends on the timing offset such that an effective channel with modest fading can exist for a specific timing offset. For this purpose, a receiver structure is proposed, which utilizes fractionally spaced samples of the symbols. In this structure, soft symbol estimates, produced by nonlinear post-distortion filters at each fraction, are combined to detect the transmitted symbols by taking the distortion for each effective channel into account.

As a side contribution, we carried out an equivalent nonlinear ISI channel model analysis, which shows that even for memoryless nonlinearity, received signal suffers from nonlinear distortion depending on the neighboring symbols due to pulse shaping. It is observed that optimal decoder requires a maximum likelihood sequence detector that maximizes PDF of the nonlinearly distorted signal. In addition, from this analysis, it can be interpreted that detectors proposed in \cite{ziya,SalehDet,hammer_twc2017} are special cases, where memory terms are not utilized in the detection. Therefore, in this study, performance improvement is achieved by including memory to the detectors even for the presence of memoryless nonlinearity.

The remainder of this paper is organized as follows. In Section \ref{Sys_model}, system model and frame structure is described and equivalent nonlinear ISI channel analysis is presented. In Section \ref{section3}, nonlinear distortion analysis is carried out for frequency selective channels and the receiver architecture is introduced. Section \ref{novel_rec} introduces novel symbol rate receiver together with proposed post-distortion algorithms. In Section \ref{perf_metrics}, performance metrics used in evaluations of the performances are presented. Simulation results are given in Section \ref{num_eval}. Lastly, concluding remarks are stated in Section \ref{concls}.

\input{system_block_diagram.tex}
\section{System Model} \label{Sys_model}
\subsection{SC-FDE Based Transmitter Model}

Considered transceiver scheme is summarized in Fig. \ref{fig:system_block}. In each block, $N_D$ number of $P$-QAM symbols, $[a_0, a_1, ..., a_{N_D -1}]^T_{N_D \times 1}$ with $\mathbb{E}\{ |a_k|^2\} = 1$, are transmitted by employing SC modulation. Cyclic prefix (CP) and cyclic suffix (CS) are added to prevent inter-block interference and create a circulant channel matrix. Then, signal vector for the transmitted block becomes, ${\bf{a}} = [a_{N_D -N_{CP}},...,a_{N_D -1},a_0, a_1, ..., a_{N_D -1},a_0,...,a_{N_{CS}-1}]^T$, where $N_{CP}$ and $N_{CS}$ are the lengths of CP and CS respectively. In discrete time, signal to be transmitted can be expressed as

\begin{equation}
x_{n}=\sum_{k= 0}^{ N-1 } {{a}}_k p_{n-\mu k} ,\label{sc_eqn}
\end{equation}
where $ p_n $ is the upsampled pulse shaping filter with upsampling factor $ \mu  $ and $N = N_D+N_{CP}+N_{CS}$. Then, the transmitted signal, ${x}_{n}=|x_n|e^{j \phi_{in}}$, is fed to PA, which nonlinearly distorts the signal. In this study, Saleh model  \cite{SalehModel}, which is commonly employed in simulations, and a realistic PA model \cite{ericsson}, which are extracted from an actual hardware are considered. For Saleh model, output of the nonlinear PA, $\tilde{x}_n= |\tilde{x}_n| e^{j\phi_{in}}e^{j\theta_n}$, can be expressed as

\begin{equation}
|\tilde{x}_n| = \frac{g_0 |{x}_n|}{1+(|{x}_n|/A_{sat})^2}, \quad \theta_n = \frac{\alpha |{x}_n|^2}{1 + \beta |{x}_n|^2}, \label{saleh_model}
\end{equation}
where $g_0$, $A_{sat}$, $\alpha$ and $\beta$ are the model parameters and they are chosen as in \cite{ziya}, $g_0=2$, $A_{sat}=1$, $\alpha=2$ and $\beta=1$ to have a realistic model. In addition to Saleh model, we considered a model that is extracted based on the measurements on the GaN power amplifier \cite{ericsson}. This model is based on memory polynomial \cite{GMP_jour} and the output of the PA is represented as
\begin{equation}
\tilde{x}_n =  \sum_{k = 0}^{K_b-1}\sum^{P_b-1}_{l=-P_b+1}\sum^{P_c-1}_{m=-P_c+1}c_{k,l,m} {x}_{n-l} |{x}_{n-l-m}|^{2k}, \label{mp_model}
\end{equation}
where $c_{k,l,m}$'s are the model coefficients. In general, PA output can represented by a generic nonlinear function as $\tilde{x}_n = \Psi(\{x_n\})$.
Then, discrete time baseband received signal model after matched filtering (MF) becomes

\begin{equation}
y_n = \left(\sum^{L-1}_{l=0}h_l \tilde{x}_{n-l} + \nu_n\right) \circledast p^*_{-n},
\end{equation}
where $h_l$'s $l=0,1,...,L-1$ are complex channel coefficients, $\circledast$ denotes convolution sum and $\nu_n$ is zero-mean complex additive white Gaussian noise (AWGN) with variance $\mathbb{E}\{ |\nu_n|^2\}=N_0$.

\input{frame_st.tex}

\subsection{Frame Structure}
The frame structure that is employed in this study is shown in Fig.  \ref{framest}. At the beginning of the transmission, a training sequence is sent to perform parameter learning for the nonlinear post-distortion. In the proposed system, nonlinearly distorted transmitted signals are captured by the observation chain of UT and sent to BS and nonlinear parameter learning (NPL) is performed under no channel impairments with high SNR at BS. Since PA characteristics varies slower compared to wireless channel, NPL, which can be referred as slow time (ST) training, is not performed very often; thus NPL does not bring any overhead to the system. ST sequence of length, $N_S$, is only transmitted when the nonlinear characteristics of the PA is changed. After NPL stage, data transmission starts. A block fading system is assumed and the channel state information (CSI) is not available. Since channel estimation is beyond the scope of this paper, a sufficiently long training sequence having $N_F$ symbols is transmitted at the beginning of each data block. Least squares (LS) method is employed to obtain CSI in fast time (FT) training stage. Details of both fast and slow time training operations will be discussed in detail in subsequent sections.



\section{ Distortion Analysis and Pre-processing : Bank of FDE} \label{section3}


In this section, analysis on the effects of nonlinear distortion on FDE performance under frequency selective channels is presented. Also, based on the analysis, a receiver architecture is proposed to decrease the effects of nonlinear distortion by exploiting channel diversity.

\subsection{Distortion Analysis for Frequency Selective Channels} \label{DistortionAnalysis}

In order to analyze the effects of frequency selective channels on nonlinearly distorted signal, signal under consideration should be decomposed into linear and nonlinear parts so that each term can be analyzed individually. Decomposition can be performed by using Bussgang decomposition as
\begin{equation}
    \tilde{x}_n = \alpha_x x_n + \gamma_n,
\quad \quad
{\alpha}_x = \frac{\mathbb{E}[x_n^*\tilde{x}_n]}{\mathbb{E}[x_n^*x_n]}, \label{buss_coef}
\end{equation}
where $\tilde{x}_n$ and $x_n$ are defined in \eqref{sc_eqn} and \eqref{saleh_model}, respectively. Bussgang coefficient, $\alpha_x$, can be found by using Wiener filtering given in \eqref{buss_coef} and $\gamma_n$ is the remaining distortion term. Received signal after matched filtering can be expressed in frequency domain, $Y(e^{j\omega})$, as
\begin{equation}
    Y(e^{j\omega}) = \alpha_x X(e^{j \omega}) H(e^{j \omega}) P^*(e^{j \omega}) + \Gamma(e^{j\omega}) H(e^{j \omega}) P^*(e^{j \omega}), \label{received_freq}
\end{equation}
where $X(e^{j \omega}) = \sum_{m= 0}^{N_D-1} a_m P(e^{j\omega})e^{-jm \omega \mu}$ is the desired signal spectrum, which is equivalent to $X(e^{j\omega}) = P(e^{j\omega}) A(e^{j \mu \omega})$  and $A(e^{j \omega}) =  \sum_{m=0}^{N_D-1} a_m e^{-jm \omega }$. In \eqref{received_freq}, $H(e^{j \omega})$ and $P^*(e^{j \omega})$ are the frequency responses of channel, $h_n$ and matched filter, $p^*_{-n}$, and the frequency response of distortion, $\Gamma(e^{j\omega}) \triangleq \sum_{n=0}^{\mu N_D-1} \gamma_n e^{-jn \omega}$, is related to power spectral density (PSD) of distortion term as, $S_\Gamma(e^{j\omega}) = \lim_{N_D \to \infty} \mathbb{E} \left [\frac{1}{\mu N_D}|\Gamma(e^{j\omega})|^2 \right]$ which is given in \cite{dinis_book} as
\begin{equation}
    S_\Gamma(e^{j\omega}) = \sum_{s=1}^{\infty} p_{\Gamma,s} \underbrace{S_X(e^{-j\omega})  \circledast  ... \circledast S_X(e^{-j\omega}) }_{s} \underbrace{S_X(e^{j\omega})  \circledast  ...  \circledast S_X(e^{j\omega}) }_{s+1}, \label{dist_spec}
\end{equation}
where  $ S_X(e^{j\omega})$ is PSD of $x_n$ and $p_{\Gamma,s}$ can be considered as the power of the $s^{th}$ order nonlinearity. Besides, it can be inferred from \eqref{dist_spec} that spectral regrowth occurs due to nonlinearity. Consequently, \eqref{received_freq} is simplified to
\begin{equation}
    Y(e^{j\omega}) = \alpha_x  A(e^{j \mu \omega}) |P(e^{j\omega})|^2 H(e^{j \omega}) + \Gamma(e^{j\omega}) H(e^{j \omega}) P^*(e^{j \omega}).
\end{equation}
After decimation, sampling rate is reduced to symbol rate, downsampled signal in frequency domain, namely $Y^d(e^{j\omega})$, can be expressed as
\begin{equation}
\begin{split}
    Y^d(e^{j\omega}) = &  \underbrace{\alpha_x A(e^{j  \omega}) \frac{1}{\mu} \sum_{i=0}^{\mu-1} H(e^{j( \frac{\omega - 2\pi i}{\mu})}) |P(e^{j(\frac{\omega - 2\pi i}{\mu})})|^2 } _{\Phi(e^{j\omega}) : \; Linear \; Term }+ \\
    & \underbrace{\frac{1}{\mu}\sum_{i=0}^{\mu-1} \Gamma(e^{j( \frac{\omega - 2\pi i}{\mu})}) H(e^{j( \frac{\omega - 2\pi i}{\mu})}) P^*(e^{j( \frac{\omega - 2\pi i}{\mu})})}_{\Psi(e^{j\omega}) : \; Distortion \; Term}, \label{eff_ch_fr}
\end{split}
\end{equation}
where $ \frac{\alpha_x}{\mu}\sum_{i=0}^{\mu-1} H(e^{j( \frac{\omega - 2\pi i}{\mu})}) |P(e^{j( \frac{\omega - 2\pi i}{\mu})})|^2$ can be interpreted as the effective channel frequency response, $H_{eff}(e^{j \omega})$. In addition, PSD of down-sampled distortion signal can be expressed as
\begin{equation}
    S_\Psi(e^{j\omega}) = \frac{1}{\mu}\sum_{i=0}^{\mu-1} S_\Gamma\left(e^{j( \frac{\omega - 2\pi i}{\mu})}\right) \left|H\left(e^{j( \frac{\omega - 2\pi i}{\mu})}\right)\right|^2 \left|P\left(e^{j( \frac{\omega - 2\pi i}{\mu})}\right)\right|^2.
\end{equation}

\begin{figure}
\centering
    \includegraphics[scale=0.6]{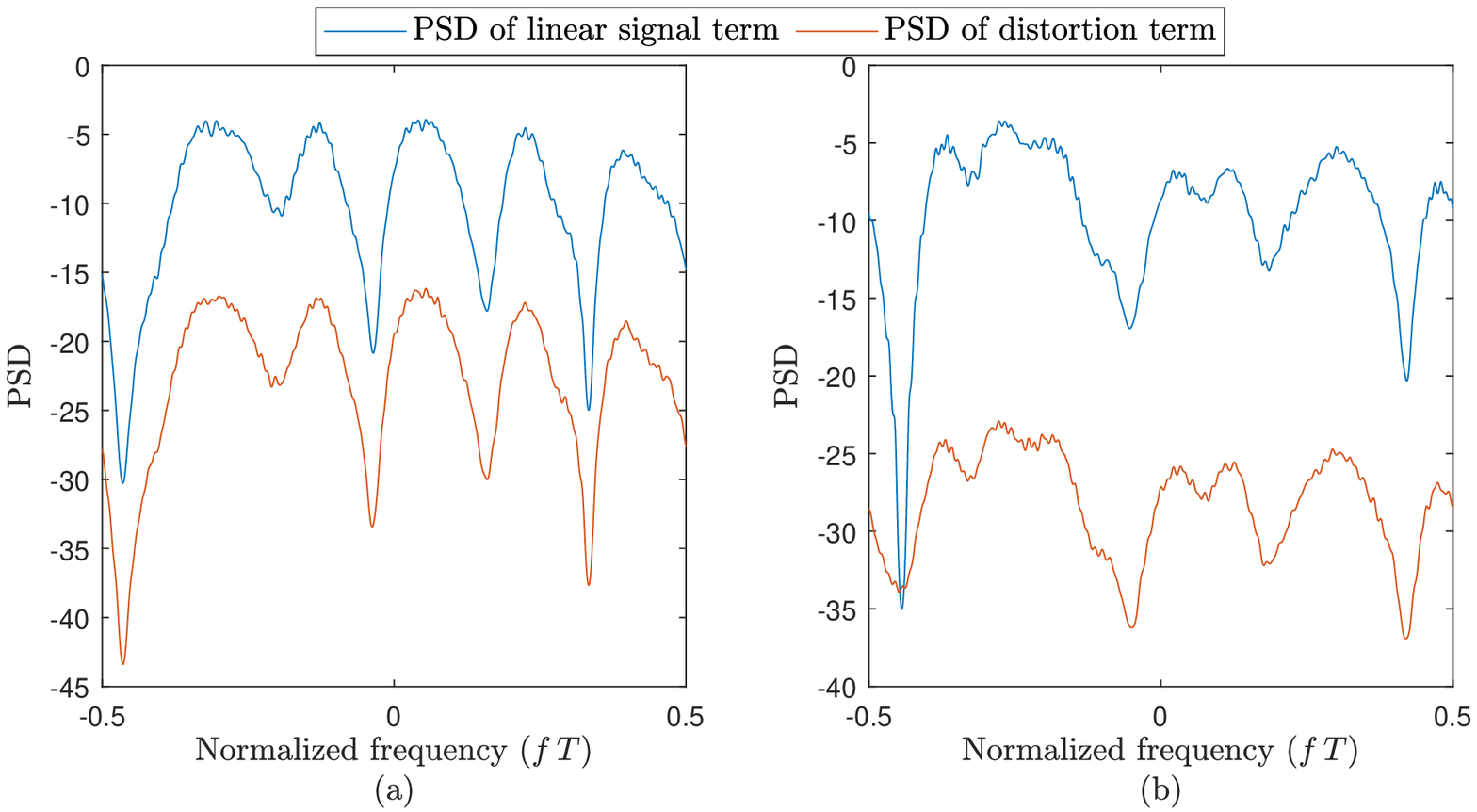}
    \caption{ PSD of signals (a) for symbol sampled channel and (b) for upsampled channel.}
    \label{equiv_ch_res}
\end{figure}

From \eqref{eff_ch_fr}, it can be observed that channels experienced by data symbols, $a_m$'s, and distortion signal are different, in general. Therefore, a particular frequency, $\omega_f$ that yields fading, $H_{eff}(e^{j \omega_f}) \approx 0$, for symbol spectrum, may not cause fading for distortion. Consequently, distortion power is amplified during FDE operation since large gain is applied to corresponding frequency component of distortion to equalize deep fades of $H_{eff}(e^{j \omega})$ even if distortion does not experience fading. On the other hand, if communication channel is sparse such that non-zero taps only exist at symbol times then $ H(e^{jw})$ becomes periodic with $\frac{2 \pi}{\mu}$. Therefore, for such channel, $H(e^{j( \frac{\omega - 2\pi i}{\mu})})$ term can be moved outside the summations. Consequently, both linear and distortion terms experience similar channels, which does not yield any distortion amplification.  In Fig. \ref{equiv_ch_res}, PSD's of both distortion and linear signal terms are shown. It is observed from Fig. \ref{equiv_ch_res} (a) that for symbol rate channel both linear and distortion terms experience the same equivalent channel. However, for a more realistic upsampled channel model, it can be observed from Fig. \ref{equiv_ch_res} (b) that linear and distortion signals are subject to different equivalent channels where linear signal is having a deep fade but distortion term is not. Consequently, distortion amplification occurs after FDE operation and amplified distortion is spread over all symbols via IDFT operation. To reduce the effects of distortion amplification, we propose a receiver structure, which is shown in Fig. \ref{bankFde}, that exploits channel diversity obtained from different sampling instances. For each sampling instance, FDE and post-distortion operations are performed and outputs for each sampling instance is combined to produce desired symbol by taking the distortion power at each branch into account.

\input{block_diagram.tex}

\subsection{Channel Acquisition}
Sampled signal is divided into $\mu$ branches, where the signal on $i^{th}$ branch can be given as

\begin{equation}
y_n^{(i)} = y_{n \mu +(i-1)}, \quad i = 1,...,\mu \label{sampling}
\end{equation}
where it can be assumed that both cyclic prefix and suffix parts of data block is discarded. For each branch standard FDE procedure is followed. Firstly, by using fast time training sequence, LS estimate of channel, ${\bf\hat{h}}^{(i)} = [\hat{h}_{-L_b+1}^{(i)},...,\hat{h}_0^{(i)},...,\hat{h}_{L_f-1}^{(i)}]^T $, for $i^{th}$ branch is obtained as

\begin{equation}
 {\bf\hat{h}}^{(i)} = ([{\bf A}]^H {\bf A})^{-1} [{\bf A}]^H {\bf y}_{fast}^{(i)}, \label{LSestimate}
\end{equation}
where  $[{\bf A}]_{k,l} = a_{(k-l)+L_b-1} $ is the data matrix for FT sequence, ${\bf A} \in \mathbb{C}^{N_D \times (L_b+L_f-1)}$, and ${\bf y}_{fast}^{(i)}  =[y_0^{(i)} , y_1^{(i)} , ..., y_{N_F -1}^{(i)} ]^T_{N_F \times 1}$ is the received sequence for FT training stage for $i^{th}$ branch. Note that estimated channel in \eqref{LSestimate}, has also anticausal parts to recover synchronization errors. 
%
  
%

\subsection{FDE Operation} FDE is performed for all branches on the received information signals. DFT of the received signal is computed as $ {\bf y}^{(i)}_f = {\bf Q}^H  {\bf y}^{(i)}$ where ${\bf y}^{(i)} = [y_{0}^{(i)}, y_{1}^{(i)}, ..., y_{N_D -1}^{(i)}]^T$ and ${\bf Q}$ is the $N_D \times N_D$ normalized DFT matrix with the $(m,n)^{th}$ element, $ q_n^m = \frac{1}{\sqrt{N_D}} e^{j2 \pi mn/N_D}$. Then, received signal in frequency domain can be expressed as
\begin{equation}
{\bf y}^{(i)}_f = \operatorname{diag} \{\lambda _{k}^{(i)} \}_{k=0}^{N_D-1} {\bf a}_f + {\bf n}^{(i)}_f,
\end{equation}
where ${\bf a}_f \triangleq {\bf Q}^H \,{\bf a}$, ${\bf n}^{(i)}_f  \triangleq  {\bf Q}^H {\boldsymbol \nu}^{(i)}$, where $[{\boldsymbol \nu}^{(i)}]_n  \triangleq (\nu_n \circledast p_{-n}^*)|_{n\rightarrow n\mu +i-1}$, and the estimated equivalent channel frequency response can be expressed as $ \lambda_{k}^{(i)} =\frac{1}{\sqrt{N_D}} \sum_{l=-L_b+1}^{L_f-1} {\hat{h}}^{(i)}_l  e^{-j2 \pi kl/N_D}$ for $k = 0,\ldots, N_D-1$. To perform FDE, minimum mean squared estimation type filtering is applied in frequency domain as
\begin{equation} 
{\bf z}^{(i)}_f = \operatorname{diag} \left\{ \frac{[\lambda_{k}^{(i)}]^*} {|\lambda_{k}^{(i)}|^2 +\delta} \right\}_{k=0}^{N_D-1} {\bf y}_f^{(i)},
\end{equation}
where $\delta = \frac{N_o}{E_s}$ is the regularization parameter, $E_s$ is the average signal power and $N_o$ is the noise variance. After performing FDE, time domain signal, $\{{z}^{(i)}_n\}$ for $i^{th}$ branch, is obtained via IDFT, ${\bf{z}}^{(i)} = {\bf Q} {\bf z}^{(i)}_f$. Following the linear equalization, nonlinear post-distortion is performed to suppress the nonlinear distortion. Details of post-distortion design will be presented in next section.

\section{A Novel Symbol Rate Nonlinear post-distortion based on FDE Bank} \label{novel_rec}

In this section, a symbol rate nonlinear post-distortion (NP) approach for higher order constellations is considered. Two different NP methods, which predicts the sequence dependent nonlinearly distorted symbols, will be presented. In addition, a decision metric, which employs FDE bank, will be introduced in order to exploit diversity to reduce distortion amplification effects.



\subsection{Post Processing: Symbol rate nonlinear post-distortion}
In this study, we consider two different symbol rate nonlinear post-distortion techniques based on GPR and augmented real-valued time-delay neural network (ARVTDNN). Aim of the proposed nonlinear post-distortion is to nonlinearly modify the incoming signal so that desired output is produced. However, one should note that we are not interested in finding the inverse model of the PA, $\Psi^{-1}(\cdot) $, which is defined in upsampled domain but performing a symbol rate nonlinear processing, which eliminates the nonlinear ISI, is proposed. A detailed analysis on equivalent nonlinear ISI channel is provided in Appendix, which shows that even if a memoryless nonlinear is considered, nonlinearity occurs due to violation of Nyquist-1 criterion. Hence, post-distortion techniques should be utilized to compensate nonlinear ISI.

\subsubsection{GPR based nonlinear post-distortion}

GPR is developed to predict output of a nonlinear system by the minimizing mean squared error in function space \cite{GPR_nl}. It can be considered as the Wiener solution of the nonlinear identification problem since system output is assumed to have Gaussian distribution. The nonlinear model based on GPR for real, $a_n^I=\operatorname{Re}\{ a_n\}$, and imaginary, $a_n^Q = \operatorname{Im}\{ a_n\}$, parts of the desired NP output is described as  \cite{Rasmussen2004,GPR_nl}
\begin{equation}
a_n^I = \Omega_I({\bf \bar{z}}_n)+\nu_n^I \quad and \quad a_n^Q = \Omega_Q({\bf \bar{z}}_n)+\nu_n^Q, \label{gpr_model}
\end{equation}
where $\nu_n^I$, $\nu_n^Q$ are the in-phase and quadrature parts of the modeling error, which are zero-mean white Gaussian processes (GP) independent of other sequences with variance $\sigma_{\nu}^2$, and the latent functions 
$\Omega_I(\cdot)$ and $\Omega_Q(\cdot)$ are the nonlinear functions to be identified. Input vector, ${\bf \bar{z}}_n$, has the augmented form as
\begin{equation}
\begin{split}
    {\bf \bar{z}}_n = [\operatorname{Re}\{ z_{n+M-1}\},...,\operatorname{Re}\{ z_{n}\},...,\operatorname{Re}\{ z_{n-M+1} \}, \\
    \operatorname{Im}\{ z_{n+M-1}\},...,\operatorname{Im}\{ z_{n}\},...,\operatorname{Im}\{ z_{n-M+1} \}& ]^T, \label{gpr_input}
\end{split}
\end{equation}
where $M$ is memory depth. Training set for GPR based post-distortion is defined as, $\{ a_n,  {\bf \bar{z}}_n \}_{n=0}^{N_s-1}$ where  $a_n$ is the desired output. In this section, we focus on modelling the in-phase component; however, one can use the same procedure for the imaginary part. In order to express distribution of GP model in \eqref{gpr_model}, required functions, vectors and matrices are defined as the followings:
\begin{itemize}
\item 
Kernel function, which is the cross correlation between different samples of the latent process, is defined as
$k({\bf \bar{z}}_p,{\bf \bar{z}}_q) \triangleq \mathbb{E} \{ \Omega_I({\bf \bar{z}}_p)\Omega_I({\bf \bar{z}}_q) \}. \label{kfunc}$
\item
Kernel matrix ${\bf K}$ is defined as the correlation matrix with entries, $[{\bf K}]_{p,q} = k({\bf \bar{z}}_p,{\bf \bar{z}}_q)$. 
\item
Kernel steering vector, namely ${\bf k}_* \triangleq \mathbb{E} \{ {\bf a}_{slow} \Omega_I({\bf \bar{z}}_*) \}$, is defined as the cross-correlation between the training and the test signals, where ${\bf a}_{slow} \triangleq [a^I_0,a^I_1,...,a^I_{N_s-1}]^T$ is the training sequence vector. It is used to predict the test symbol, $a^I_*$ by using $z_*$, where $*$ denotes time indices in data sequence, $p^{th}$ element of ${\bf k}_*$ can be calculated as $[{\bf k}_*]_p = k({\bf \bar{z}}_p,{\bf \bar{z}}_*)$ due to the independence of $\Omega_I({\bf \bar{z}}_p)$ and $\nu_p^I$. 
\end{itemize} 
 
Based on defined kernel functions, the predictive distribution can be written as \cite{Rasmussen2004,GPR_nl}
\begin{equation}
    p({ {a}^I_*} | \{ a_n,  {\bf \bar{z}}_n \}_{n=0}^{N_s-1}, {\bf \bar{z}}_*) \sim  \mathcal{N} \left( \mu_*,\sigma^2_* \right), \label{predictPdf}
\end{equation}
where $\mu_*$ and $\sigma^2_*$ are mean and variance of the distribution, and they can be obtained as
\begin{equation}
    \mu_* = {\bf k}_*^T \left( {\bf K} + \sigma_{\nu}^2 {\bf I}\right)^{-1}{\bf a}_{slow}, \quad
    \sigma^2_* = k({\bf \bar{z}}_*,{\bf \bar{z}}_*) - {\bf k}_*^T ({\bf K} + \sigma_{\nu}^2 {\bf I})^{-1} {\bf k}_* + \sigma_{\nu}^2. \label{predict_var}
\end{equation}
Then, symbol estimates at the output of the nonlinear post-distortion becomes the mean of the predictive distribution:
\begin{equation}
    \tilde{a}_*^I = \mathbb{E}\{ a_*^I| \{ a_n,  {\bf \bar{z}}_n \}_{n=0}^{N_s-1}, {\bf \bar{z}}_* \} = {\bf k}_*^T \left( {\bf K} + \sigma_{\nu}^2 {\bf I}\right)^{-1}{\bf a}_{slow}.
\end{equation} 
In this study, kernel function is chosen as the exponential function that is widely employed in literature, $k({\bf \bar{z}}_p,{\bf \bar{z}}_q) = \sigma_f^2 \operatorname{exp}\left( -{\bf \bar{z}}_p^T {\bf Q}^{-1}{\bf \bar{z}}_q\right)$, where ${\bf Q}=\operatorname{diag}\{c_i^2\}^{4M-2}_{i=1}$ and $c_i$'s are the length scale parameters and $\sigma_f^2$ is the signal variance. Maximum likelihood estimation is used to find the hyper-parameters, $\beta = [\sigma_f,\sigma_\nu, c_1, ..., c_{4M-2}]^T$ \cite{Rasmussen2004,post1}.

Hyper-parameters optimization is a time consuming process and training time exponentially increases with the length of the training sequence. Hence, higher order $P$-QAM processing requires a long training sequence to train GPR parameters. In order to reduce the training time, we propose BLUE type post processing such that training sequence is divided into non-overlapping segments to train different GPR units and intelligently combine the output of each unit.

Consider, estimated symbol for $i^{th}$ segment with training set $\{ a_n^{(i)}, {\bf \bar{z}}_n^{(i)} \}_{n=0}^{N_S^{(i)}-1}$, $ \tilde{a}^{i}_n $, which is obtained as $\tilde{a}_*^{I,(i)}  =   [{\bf k}^{(i)}_*]^T \left( {\bf K}^{(i)} + \sigma_{\nu}^2 {\bf I}\right)^{-1} {\bf {a}}_{slow}^{(i)} $, where ${\bf k}^{(i)}_* $ is the cross-correlation vector, ${\bf K}^{(i)} $ is the kernel matix, and ${\bf {a}}_{slow}^{(i)}$ is the training symbols for $i^{th}$ segment, $N_S^{(i)}$ is the training length for $i^{th}$ segment, and $S$ is the number of segments: $\sum_{i=1}^{S} N_S^{(i)} = N_S$. For simplicity, we dropped I/Q indices. Also, variance, $({\sigma_*^{i})^2} $, of each symbol estimate can be obtained as
\begin{equation}
    ({\sigma_*^{i})^2} = k({\bf \bar{z}}^{(i)}_*,{\bf \bar{z}}^{(i)}_*) -  [{\bf k}^{(i)}_*]^T \left( {\bf K}^{(i)} + \sigma_{\nu}^2 {\bf I}\right)^{-1} [{\bf k}^{(i)}_*] + \sigma_{\nu}^2,
\end{equation}
where ${\bf \bar{z}}^{(i)}_*$ is test data for $i^{th}$ segment. Estimates are fused based on BLUE combiner as in \cite{gokhan_hoca_blue}
\begin{equation}
    \tilde{a}_* = \sum_{i=1}^{S} w^{i}_* \tilde{a}^{(i)}_*,
\end{equation}
where ${\bf w}_* = [ w_*^{1}, ..., w_*^{S}]^T$ is the weights, which are obtained as, ${\bf w}_* = \frac{{\bf D}_*^{-1} {\bf 1}}{{\bf 1}^T {\bf D}_*^{-1} {\bf 1} }$, where ${\bf D_*} \triangleq \operatorname{diag}\{({\sigma_*^{1})^2},...,({\sigma_*^{S})^2}\}$ is the diagonal variance matrix.


\subsubsection{Neural Network (NN) based nonlinear post-distortion}

In this study, ARVTDNN structure shown in Fig. \ref{fig:ARVTDNN}, which is employed to design digital predistortion unit in \cite{arvtdnn} via behavioral modelling, is used as the nonlinear post-distorter. However, functionality of the NN in this study is different compared to that of DPD. In \cite{arvtdnn}, NN is used for nonlinear system identification via nonlinear regression so that inverse of the nonlinear function is obtained. However, in this study, we are employing NN to estimate transmitted symbols chosen from a discrete alphabet. 

\input{NNArch.tex}
The same input signal, ${\bf \Bar{z}}_n$, which is defined for GPR estimation, is also used for NN post-distortion. Then, symbol estimate at the output of NN can be expressed as $ \tilde{a}_n = \Omega_I({\bf \Bar{z}}_n) + j\Omega_Q({\bf \Bar{z}}_n)$, where $\Omega_I({\bf \Bar{z}}_n)$ and $\Omega_Q({\bf \Bar{z}}_n)$ are the estimates of real and imaginary parts respectively.
The cost function is defined as
\begin{equation}
    J = \frac{1}{2N_S} \sum_{n=0}^{N_S -1}  (\operatorname{Re}\{a_n\} -\Omega_I({\bf \Bar{z}}_n))^2 + (\operatorname{Im}\{a_n\} -\Omega_Q({\bf \Bar{z}}_n))^2. \label{costfn}
\end{equation}
In the proposed network structure, there is a single hidden layer together with one input and output layers. Therefore, network output can be expressed as
\begin{equation}
    \Omega_I({\bf \Bar{z}}_n) = {\bf w}_{I}^T g({\bf W_1}{\bf \Bar{z}}_n + {\bf b}_1) + { b}_{2,I}, \label{NNexpression}
\end{equation}
where ${\bf w}_{I} \in \mathbb{C} ^{L_1 \times 1}$ is the weights for the output layer, similarly output for the quadrature part is obtained by the weight vector $ {\bf w}_{Q}$ and $b_{2,Q}$. ${\bf W_1} \in \mathbb{C} ^{L_1 \times 4M-2}$ is the weight matrix for the input layer with elements $[{\bf W_1}]_{k,l} = w_{k,l}$ which connects to $l^{th}$ neuron of the input layer to $k^{th}$ neuron at the hidden layer, ${\bf b}_1$ and ${\bf b}_2$ are the bias vectors, $L_1$ is number of neurons in hidden layer and $g(\cdot )$ is the activation function. In this study, hyperbolic tangent sigmoid transfer function, $g(x) = \frac{2}{1+e^{-2x}}-1$, is employed, which introduces the nonlinearity required for nonlinear compensation.At each epoch, cost function is evaluated and weights of NN coefficients are updated via back-propagation by using the Levenberg–Marquardt algorithm \cite{levenberg_marq}.

\subsection{An Adaptive QAM Detection based on nonlinear post-distortion bank}

In this section, a novel detector structure is proposed to compensate the distortion amplification during FDE. In order to reduce effects of distortion amplification, use of channel diversity is proposed in this study. Effective channel in symbol sampled domain varies with fractional delays of the sampling instant, which may not have deep fades in frequency domain. Therefore, we propose that fusion of the estimates for different sampling instances improves the overall detection performance by taking the distortion powers for different effective channels into account.

In Fig. \ref{bankFde}, nonlinear post-distortion produces soft symbol estimates, ${ \tilde{a}}_n^{(i)} = \Omega\left(\left\{{z}^{(i)}_n\right\}\right)$ as output for each branch, where $\Omega(\cdot)$ denotes the output of nonlinear regression function. Then soft symbol estimates, ${\bf \tilde{a}}_n = [{ \tilde{a}}_n^{(1)},{ \tilde{a}}_n^{(2)},...,{ \tilde{a}}_n^{(\mu)}]^T$, are fed into the distortion-aware symbol by symbol detector (DA-SSD), which is based on the Bussgang decomposition
\begin{equation}
    {\bf \tilde{a}}_n = {\boldsymbol \beta} a_n + {\boldsymbol \eta}, \label{vector_buss}
\end{equation}
where ${\boldsymbol \beta}$ is the vector Bussgang coefficient vector computed at fast time training period as
\begin{equation}
    {\boldsymbol \beta} =  \frac{\mathbb{E}\{ {\bf \tilde{a}}_n a_n^* \}}{ \mathbb{E}\{ |a_n|^2 \}} \approx \left(\sum_{n = 0}^{N_F -1}  {\bf \tilde{a}}_n a_n^* \right) \left(\sum_{n = 0}^{N_F -1}  |a_n|^2 \right)^{-1},
\end{equation}
and ${\bf R}_{\eta}$ is the autocorrelation matrix of distortion vector, namely ${\boldsymbol \eta}$, can be calculated as
\begin{equation}
    \begin{split}
    {\bf R}_{\eta} = { \mathbb{E}\{ ({\bf \tilde{a}}_n - {\boldsymbol \beta} a_n) ({\bf \tilde{a}}_n - {\boldsymbol \beta} a_n) ^H\}} 
    \approx \frac{1}{N_F - 1} \sum_{n = 0}^{N_F -1} ({\bf \tilde{a}}_n - {\boldsymbol \beta} a_n) ({\bf \tilde{a}}_n - {\boldsymbol \beta} a_n)^H. \label{acm_dist}
    \end{split}
\end{equation}
%
Consequently, assuming Gaussian distribution, PDF of the received signal vector, ${\bf \tilde{a}}_n$, can be expressed in terms of FDE bank outputs as
\begin{equation}
\tilde {p}({\bf \tilde{a}}_n|a_n)  = \frac{1}{(\pi )^{\mu} |{\bf R}_{\eta}|} \operatorname{exp}\left[ - ({\bf \tilde{a}}_n - {\boldsymbol \beta} a_n)^H {\bf R}_{\eta}^{-1} ({\bf \tilde{a}}_n - {\boldsymbol \beta} a_n) \right]. \label{bank_PDF}
\end{equation}
By using the PDF in \eqref{bank_PDF}, distortion-aware symbol by symbol detector can be obtained as,
\begin{equation}
    {\hat{a}}_n = \arg\min_{a_n} \quad ({\bf \tilde{a}}_n - {\boldsymbol \beta} a_n)^H {\bf R}_{\eta}^{-1} ({\bf \tilde{a}}_n - {\boldsymbol \beta} a_n). \label{da_ssd}
\end{equation}

Detector expressed in \eqref{da_ssd} can also be interpreted as maximum ratio combiner (MRC) after whitening filter since it combines different branch outputs by considering their error covariances, where higher weights are assigned to the branch outputs, which are subjected to less distortion.

\section{Achievable Information Rate (AIR) based on mismatched decoding capacity} \label{perf_metrics}


In order to evaluate achievable information rate (AIR) performance of the proposed receiver, a lower bound on the constraint capacity in terms of GMI metric is considered. In \cite{GMI:1,GMI:2,GMI:3}, mismatched decoding capacity is given as
\begin{equation}
C_P =\log_2P - E_{a,\tilde{\bf{a}}}\left[\log_2\left(\frac{\sum_{a'\in A_a} \tilde{p}(\tilde{\bf{a}}|a')}{\tilde{p}(\tilde{\bf{a}}|a)}\right)\right], \label{mis_c}
\end{equation}
where $A_a$ is QAM symbol alphabet, $P$ is the modulation order. $ \tilde {p}(\tilde{\bf{a}}|a') $ is the mismatched probability density function (PDF) that is used since exact knowledge on PDF is not available due to nonlinearity. If the channel under consideration were linear AWGN channel then $\tilde {p}(\tilde{\bf{a}}|a')$ would have the Gaussian form. However, due to lack of knowledge, an approximate exponential PDF, $ \tilde {p}(\tilde{\bf{a}}|a) $, is assumed based on Bussgang decompostion given in \eqref{vector_buss}. Therefore, in order to obtain AIR bound for the proposed receiver structure, mismatched PDF expression in \eqref{mis_c}, should be substituted by the PDF given in \eqref{bank_PDF}.

\emph{Outage probability} is another performance metric that can be used to measure the performance of the detectors in fading channels. Outage probability, $P_{out}$ can be defined as the probability of the instantaneous capacity being less then a threshold capacity $C_P^{T}$ : $P_{out} = \operatorname{Pr}\{ C_P < C_P^{T}\}$. It can also be inferred as the packet error rate since significant block errors occur if instantaneous capacity of the system falls below of the defined threshold.


\section{Numerical Evaluations} \label{num_eval}

In this section, numerical results are presented in order to evaluate the performance of the proposed decoder structure. Proposed nonlinear distortion methods are compared with the state-of-the-art methods, which can be referred as modified metric (MM), presented in \cite{ziya} and \cite{SalehDet} and Volterra series (VS) \cite{GMP_jour} based nonlinear post-distorter. Two different scenarios employing higher order modulation schemes are considered for performance evaluations.

In the first scenario, Saleh model is used as memoryless nonlinear PA and AWGN channel with $1024$ $QAM$ constellation  is considered. For this model, input signal power is scaled such that normalized output power, which is the ratio of average output power to the maximum output power of the amplifier, refers to output backoff. For nonlinear parameter learning, a slow time training sequence of length $N_S = 16384$ symbols is used.  

In the second scenario, model extracted from an actual GaN PA \cite{ericsson} and another PA available at \cite{chalmers} is utilized. In order to test the performance of the proposed architecture, transmitted signal is passed through a dispersive channel generated according to Rayleigh distributed COST-207 channel model \cite{cost207}. In simulations, channel length is chosen to cover $16$ symbols with unity gain. Also, constellation order is selected as $256$ $QAM$. In order to perform FDE operation, a channel acquisition procedure is performed by using $N_F = 3000$ training symbols. A slow time training sequence of length $N_S = 16384$ symbols is used for nonlinear parameter learning. In data transmission, block length is selected to be $N_D = 8192$ symbols. A \emph{root raised cosine} filter with $0.3$ roll-off factor used as the pulse shaping filter.

In order to utilize proposed BLUE combiner for GPR estimates, training sequence is divided into $8$ segments each having $2048$ symbols so that $8$ GPR estimates are fused. An ARVTDNN with a hidden layer having 30 neurons is employed for both scenarios. Memory depth for GPR, NN and Volterra series with linear-cubic model \cite{GMP_jour} based post-distorters is chosen as $\pm 2$ samples.

\begin{figure}
\centering
    \includegraphics[scale=0.7]{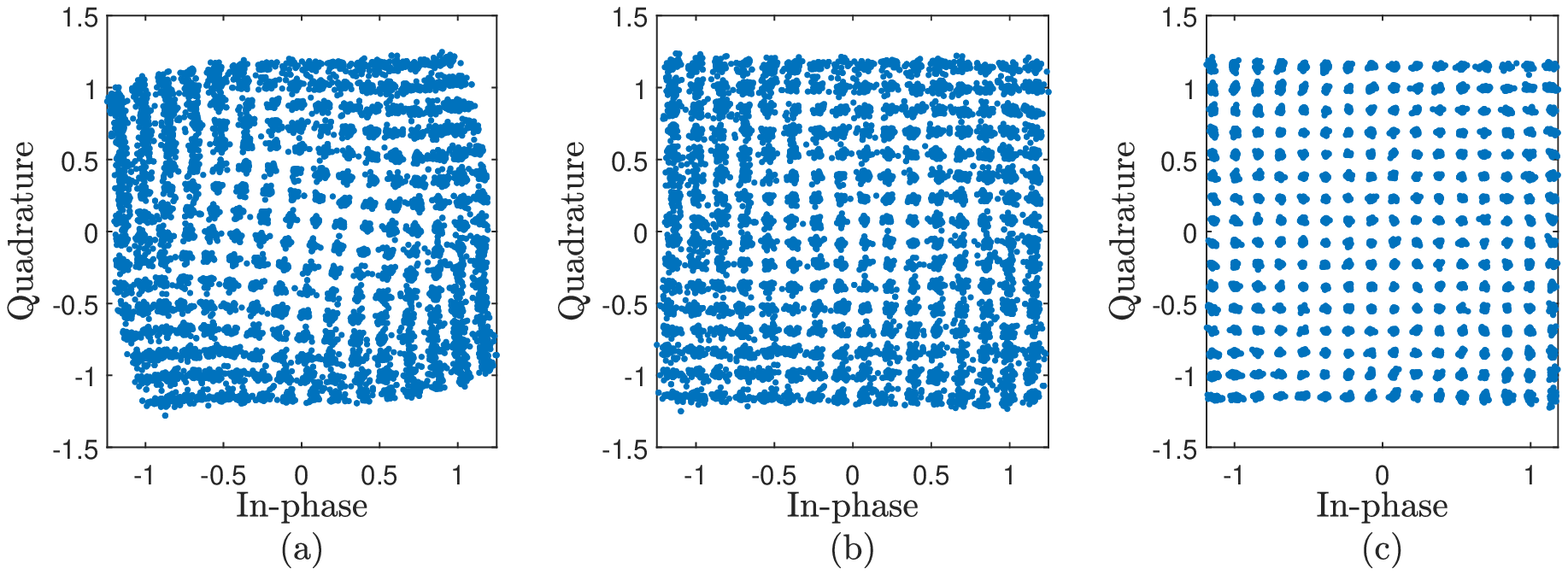}
    \caption{Scatterplots for different receivers (a) conventional recevier (b) receiver with MM (c) receiver with post distortion with NN.}
    \label{scatterplot}
\end{figure}

Before investigating the performances of the receiver structures in detail, received symbols are shown in Fig. \ref{scatterplot} for Saleh model with $5.93$ $dB$ output backoff. In Fig. \ref{scatterplot} (a), warping effect due to nonlinear behavior can be observed since conventional receiver do not attempt to compensate nonlinearity. Receiver with MM performs a memoryless constellation point dependent correction as shown in Fig. \ref{scatterplot} (b). However, there is a strong nonlinear ISI remaining. Post-distortion methods compensates nonlinear ISI together with the warping effect as can be seen in Fig. \ref{scatterplot} (c).

\subsection{Performance evaluations for AWGN channel with Saleh Model PA}

\begin{figure}
\centering
    \includegraphics[scale=0.6]{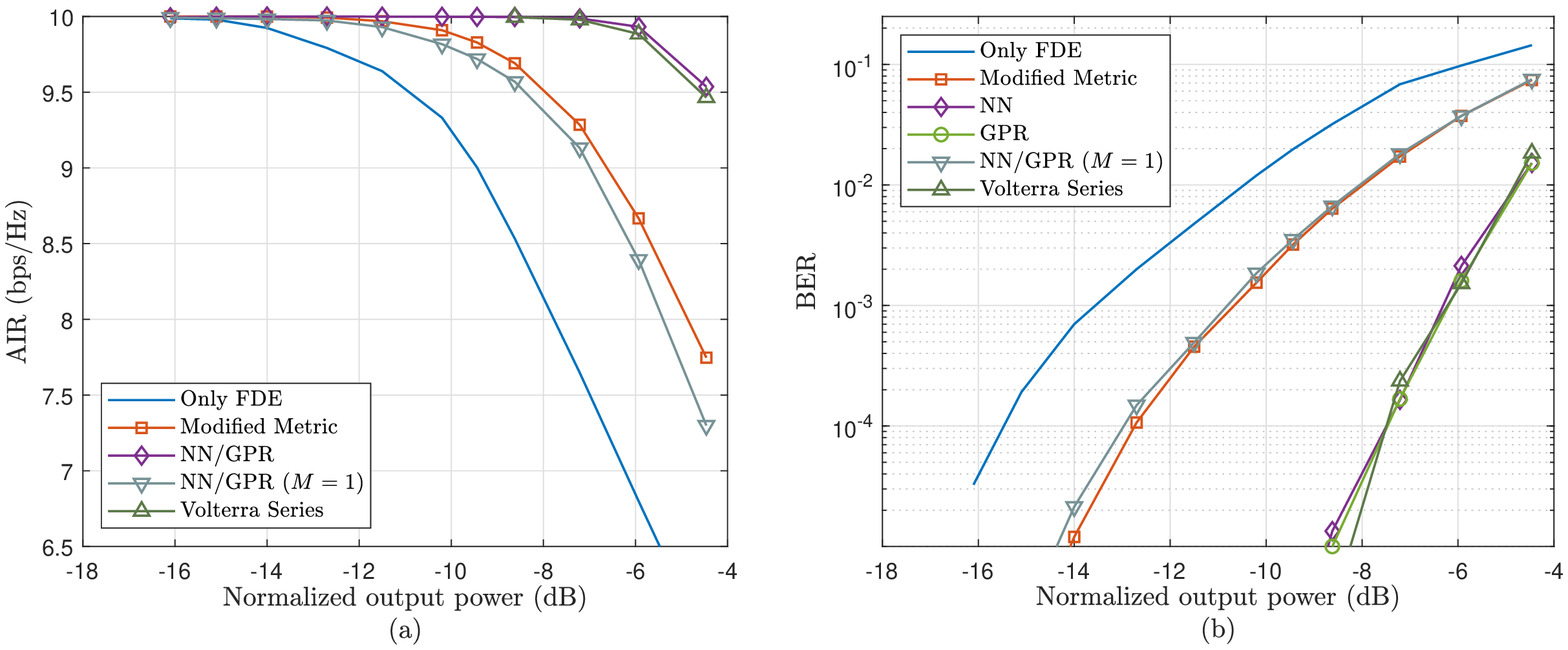}
    \caption{AIR (a) and BER (b) of the receivers for Saleh Model for different output backoffs.}
    \label{saleh_air_ber_backoff}
\end{figure}

Firstly, achievable information rate analysis is carried out for different output backoff values. For this scenario, bank of FDE is not utilized since AWGN channel with $E_s/N_o = 50$ dB is considered, where fractional delay is not an issue. In order to observe the effects of memory, memoryless post-distortion, with memory depth $M=1$, is also considered. In Fig. \ref{saleh_air_ber_backoff}(a), AIR's, obtained by GMI analysis, is shown for different receiver structures. It can be observed that even for a low backoff level, such as $4.46$ $dB$, significantly higher information rate, $9.5$ bps/Hz, is achieved by post-distortion methods, employing NN, GPR and Volterra series, compared to the conventional and MM based receivers. Lastly, need of the involving memory in detection is obvious since performance of memoryless post-distrtotion yields the same performance with MM detector, which is in compliance with nonlinear ISI channel analysis given in the Appendix.

In addition to AIR analysis, uncoded BER performances of these receivers are compared. From Fig. \ref{saleh_air_ber_backoff}(b), it can be seen that proposed NN/GPR with memory and Volterra series post-distorters significantly outperform the other methods. Similar to AIR analysis, MM provides superior performance compared to conventional receiver. Quantitatively, receivers employing post-distorters achieve $10^{-4}$ BER at $\sim 7.5$ $dB$ backoff while MM can achieve the same performance at $\sim 12.5$ $dB$ and conventional receiver achieves at $\sim 15$ $dB$ backoff.


Robustness of the proposed receivers against the noise is an important measure since all operations on the receiver side are performed on the signals, which are corrupted by the AWGN. For this purpose, AWGN is added to the received signal and $E_s/N_o$, which is the received SNR, is adjusted by modifying the receiver noise power. It should be noted that a system with linear PA is also considered as the benchmark for the performance.
\begin{figure}
\centering
    \includegraphics[scale=0.6]{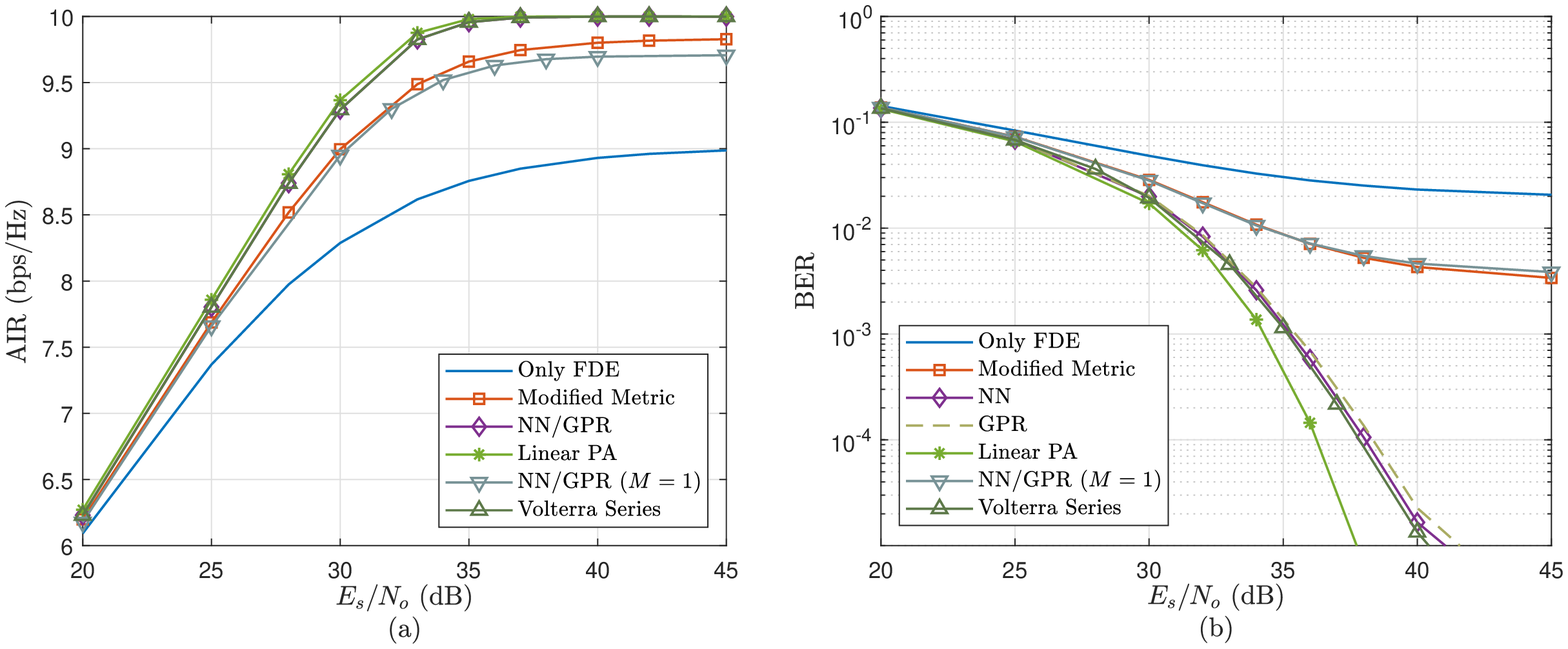}
    \caption{ AIR (a) and BER (b) vs. SNR curves for the receivers for Saleh Model where 9.44 dB output backoff is employed.}
    \label{saleh_air_snr}
\end{figure}
In Fig. \ref{saleh_air_snr}(a), achievable rates for different SNR levels are presented for $9.44$ $dB$ output backoff. Post-distortion methods based on NN, GPR and Volterra series, perform very close to ideal. However, achievable rate of MM is saturated around $9.8$ bps/Hz. Besides, conventional receiver suffers from performance degradation since its AIR saturates around $9$ bps/Hz. From Fig. \ref{saleh_air_snr}(b), it can be seen that superiority of the post-distortion methods are apparent, whereas conventional and MM based receivers suffer from high error floor.
\subsection{Performance evaluations for GaN PA Model}

\begin{figure}
\centering
    \includegraphics[scale=0.6]{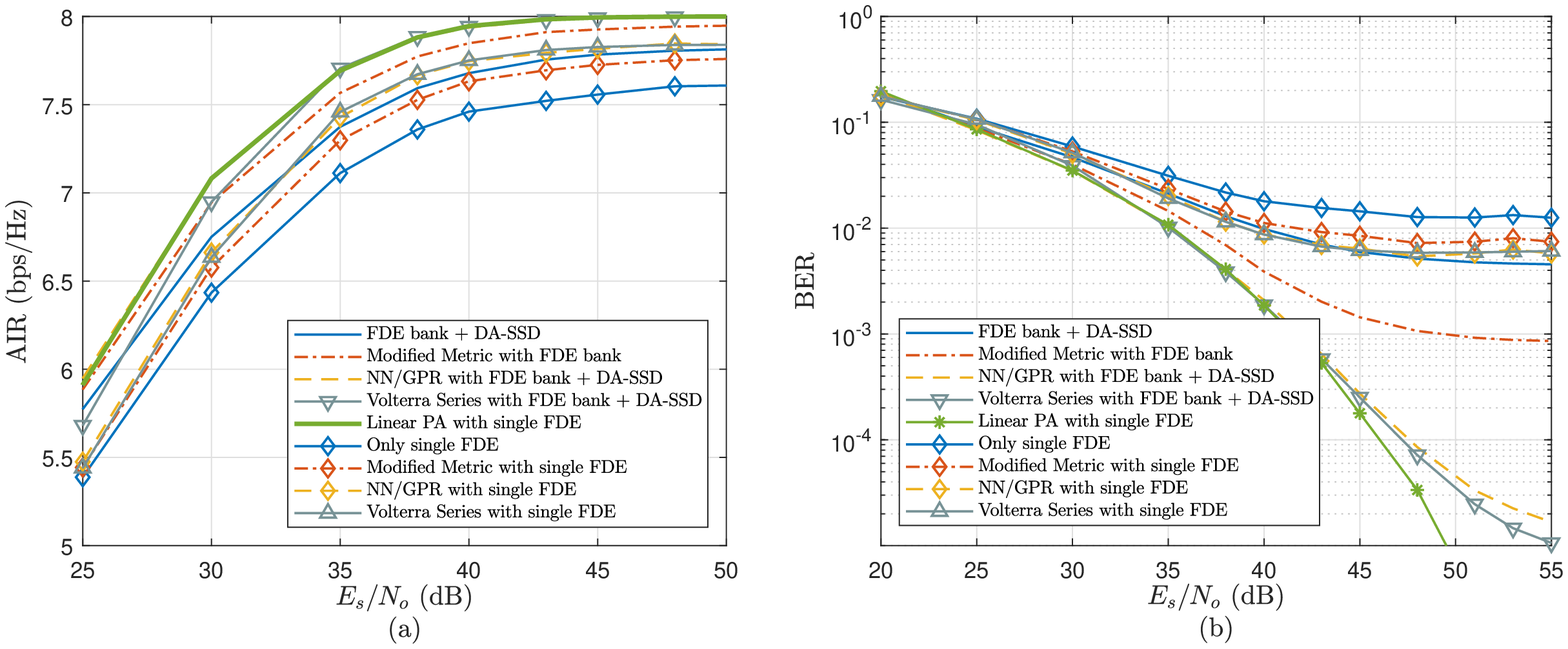}
    \caption{ AIR (a) and BER (b) vs. SNR curves for the receivers for GaN Model \cite{ericsson} .}
    \label{eric_air_ber}
\end{figure}

In this section, simulation results for more realistic hardware models and communication medium are presented for $256$ $QAM$. Firstly, performances of receivers are evaluated for PA model given in \cite{ericsson}. In Fig. \ref{eric_air_ber}(a), AIR performances are shown for different receiver architectures. It can be observed that, proposed DA-SSD improves the performance of the post-distortion schemes significantly. Besides, for MM based detectors, fractionally delayed FDE bank is also utilized and branch, which provides the highest capacity during fast time training, is chosen to decode information sequence. Selection of the branch with maximum AIR also improves the performance of the corresponding detectors. Average AIR's of the receivers using single FDE are saturated before reaching the maximum constrained capacity even if post-distortion is performed. However, post-distortion algorithms together with DA-SSD are able to attain the maximum capacity.

In Fig. \ref{eric_air_ber}(b), BER performances of the receivers are evaluated. Similar to AIR analysis, detectors that rely on a single FDE suffer from significant error floor. However, employing fractionally delayed FDE bank with DA-SSD decreases the error floor levels substantially. It is observed that post-distortion methods employing DA-SSD performs very close to linear PA. In addition, performance of memoryless MM based detector is far-from that of the post-distortion based detectors. However, it still provides improvement compared to the conventional detector.

\begin{figure}
\centering
    \includegraphics[scale=0.6]{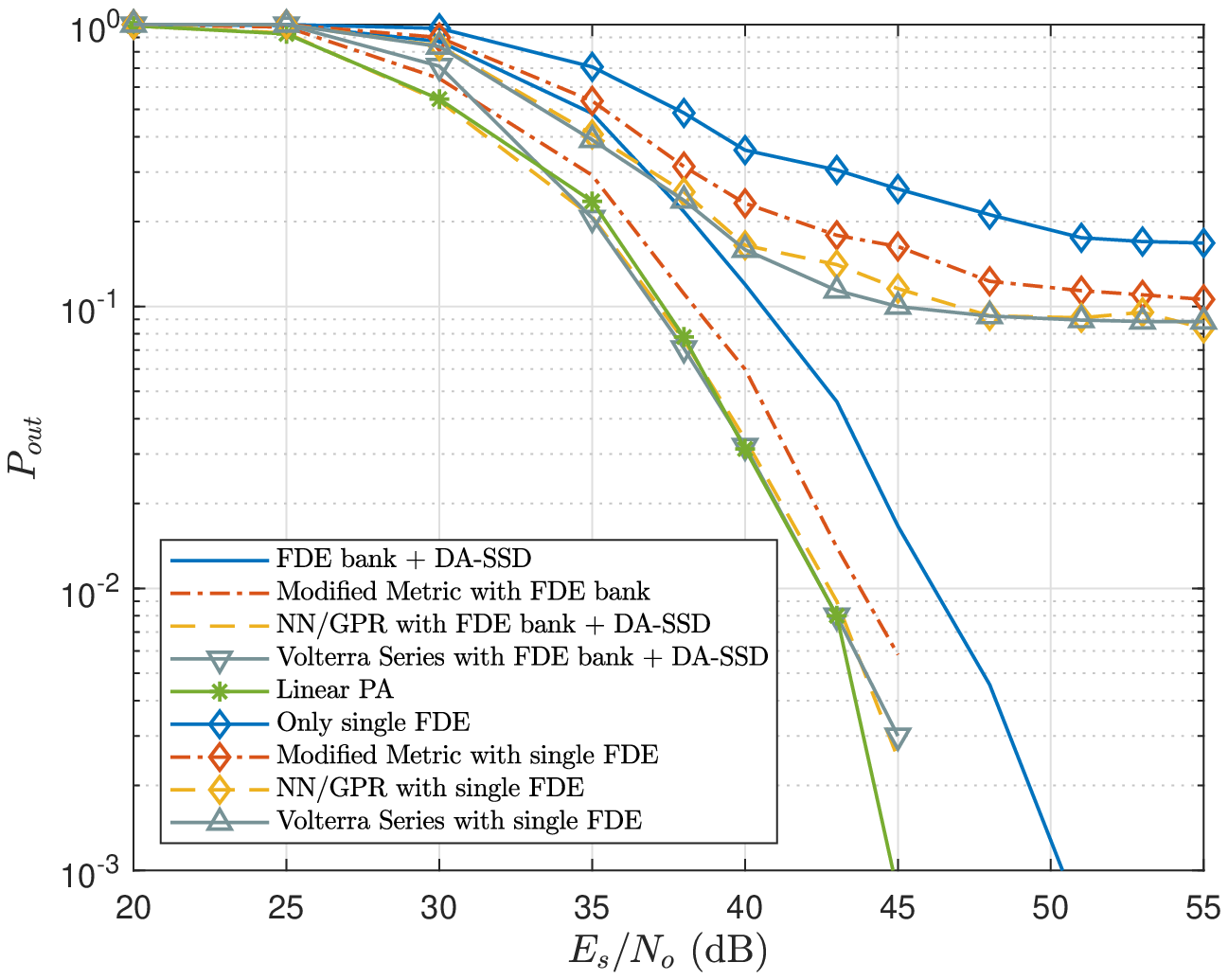}
    \caption{ Outage curves of the receivers for GaN Model \cite{ericsson}.}
    \label{eric_out}
\end{figure}

In order to evaluate the block error performance of the detectors, outage probabilities are presented in Fig. \ref{eric_out} for different SNR values. In simulations, threshold for the capacity is chosen as $C_P^T=7.5$. It is apparent that receivers with single FDE branch experience outage very often; hence, it can be concluded that observed AIR and BER performance degradation stem from fluctuations in the capacity of the system  due to channel variations. When the capacity falls below the threshold, block errors occur and overall BER increases. On the other hand, DA-SSD with FDE bank reduce outage probability by giving more weight to the branch with effective channel having modest fading and distortion.

\begin{figure}
\centering
    \includegraphics[scale=0.6]{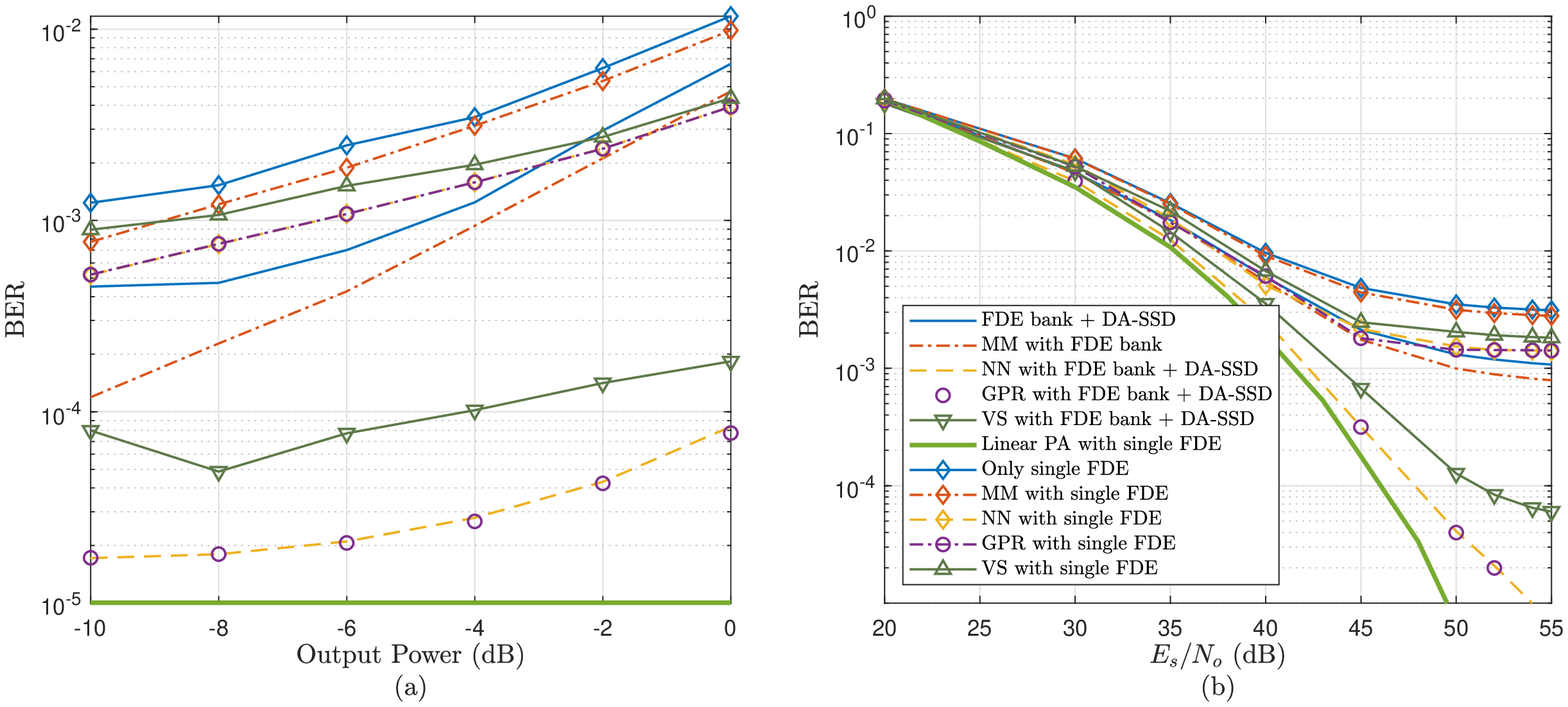}
    \caption{ BER curves of the receivers for GaN PA \cite{chalmers} for (a) different input backoffs  (b) SNR levels.}
    \label{hw_ber_backoff}
\end{figure}

Lastly, BER performances of the receivers are compared for an actual hardware given in \cite{chalmers}. BER results are shown for different output powers in Fig. \ref{hw_ber_backoff}(a), where output power is normalized such that maximum output power is scaled to be $0$ $dB$. For the simulations, noise level is set to $E_s/N_o = 50 dB$ so that $10^{-5}$ BER is achieved by the ideal hardware. It can be observed that performance of NN/GPR based post-distortion methods approaches that of linear PA. On the other hand, VS based post-distortion suffers from performance degradation compared to NN/GPR post-distorters while providing improvements in comparison to memoryless detectors.

In Fig. \ref{hw_ber_backoff}(b), BER curves for different SNR levels are demonstrated for output power $-4$ $dBm$. It is observed that BER performance of the proposed NN/GPR with DA-SSD detector performs close to ideal system while Volterra series based nonlinear distorter cannot prevent error floor. Similar to previous cases, receivers, which do not employ FDE bank, suffer from significant performance loss compared to receivers employing FDE bank. 
\section{Conclusions} \label{concls}
In this paper, we addressed the effects of PA nonlinearities impaired with memory. It is shown that even if a memoryless nonlinearity exists, nonlinear ISI occurs since Nyquist-1 criterion is violated. Therefore, we developed a post-distortion scheme in order to reduce nonlinear distortion power by taking the nonlinear ISI into account. Proposed scheme divides the problem into two independent parts. In the first part, standard FDE procedure is employed, where linear wireless channel is equalized so that dimension of the problem is reduced. Then, it is followed by the nonlinear processing, which suppresses the nonlinear ISI. In addition, effects of dispersive communication medium is considered. For this case, it is observed that FDE performance is significantly affected by the nonlinear PA due to distortion amplification. In order to overcome this problem, we propose to utilize samples obtained by fractional delayed sampling so that channel diversity is obtained. Channels obtained by using samples with different timing offsets may have different fading characteristics, which can be utilized in detection. By taking the fading effects of each timing offset into account, a novel DA-SSD based on FDE bank, which combines different symbol estimates obtained from fractionally delayed samples of the received signal, is proposed. Performance improvement provided by the proposed structure is shown via numerical simulations. It is observed that proposed post-distortion significantly outperforms the detectors that does not take the memory effects into account. In addition, it is shown that if the memory of the proposed structure is discarded, its performance converges to memoryless state-of-the-art detectors. Simulations are also performed for the scenarios where dispersive communication channels are present. Numerical results show that proposed DA-SSD significantly improves the performance of the receiver compared to receiver, which does not utilize FDE bank, and its performance is very close to that of linear system.
%


\ifCLASSOPTIONcaptionsoff
  \newpage
\fi

\appendix[Discrete-Time Equivalent Model for Nonlinear ISI Channel] 

In order to gain insight into the characteristics of nonlinear distortion, distortion term in the received signal is analyzed. For analytical convenience, we considered a memoryless nonlinear model, expressed by arbitrary nonlinear basis functions, $\psi_q(\cdot)$, where $q$ is the nonlinearity order. Then, we can express transmitted signal after nonlinear amplification as,
\begin{equation}
    \tilde{x}_n = \sum_{q=0}^{Q-1} \varpi_q \psi_q \left(\sum_{k= 0}^{ N-1 } {{a}}_k p_{n-\mu k} \right)
\end{equation}
where $\{\varpi_q\}$ are the model coefficients and $\psi_0(x)= x$ corresponds to linear term. For simplicity, noise-free line-of-sight channel is considered. After MF, the received signal can be written as,

%
\begin{equation}
    y_n = \varpi_0 \sum_{k= 0}^{ N-1 } {{a}}_k\hat{p}_{n-\mu k} + \sum_{l=-\infty}^{\infty} \sum_{q=1}^{Q-1} \varpi_q \psi_q \left(\sum_{k= 0}^{ N-1 } {{a}}_k p_{l-\mu k}\right)p^*_{n-l},
\end{equation}
where $\hat{p}= {p}_{n}\circledast p^*_{-n}$ is ideal Nyquist-1 raised cosine impulse response. After sampling, $n = \mu m$, received signal at symbol rate becomes,
%
%
%
\begin{equation}
    y_m = \varpi_0 a_m +  \sum_{q=1}^{Q-1} \sum_{l=-\infty}^{\infty}  \varpi_q \psi_q\left(\sum_{k= 0}^{ N-1 } {{a}}_k p_{l-\mu k}\right)p^*_{\mu m-l}. \label{sampNonlin}
\end{equation}
From \eqref{sampNonlin}, it can be observed that the term in $\psi_q (\cdot)$ is a unique function for a specific sequence, $\{ a_n\}$. Consequently, $\sum_{l=-\infty}^{\infty}  \psi_q\left(\sum_{k= 0}^{ N-1 } {{a}}_k p_{l-\mu k}\right)p^*_{\mu m-l}$ can be expressed by a nonlinear mapping which depends on symbol index $m$ and symbol sequence, $\{ a_n \}$. Therefore, \eqref{sampNonlin} can also be represented by employing a basis change as,
\begin{equation}
     y_m = \varpi_0 a_m +  \sum_{q=1}^{Q-1}   \varpi_q \psi^{(\{a_n\},m)}_q\left( \sum_{l=-\infty}^{\infty} \sum_{k= 0}^{ N-1 } {{a}}_k p_{l-\mu k} (\tilde{p}_{\mu m-l}^{(q)})^*\right), \label{modifBasis}
\end{equation}
where $\psi^{(\{a_n\},m)}_q (\cdot)$ represents the unique mapping, which depends on the sequence $\{a_n\}$ and sampling instant $m$. $\tilde{p}_{l}^{(q)}$ is the equivalent matched filter impulse response distorted by the nonlinearity, $\psi_q(\cdot)$, which is also defined by another sequence dependent nonlinear mapping\footnote{In this work, we are not interested in finding the mapping, but existence of such mapping is sufficient for rest of the analysis.}. By changing the summation order in nonlinearity, the expression is further simplified to

\begin{equation}
     y_m = \varpi_0 a_m +  \sum_{q=1}^{Q-1}   \varpi_q \psi^{(\{a_n\},m)}_q\left(  \sum_{k= 0}^{ N-1 } {{a}}_k  \bar{p}_{m-k}^{(q)}  \right). \label{distMF}
\end{equation}
If impulse response, $\bar{p}^{(q)}_{m} ={p}_{l}\circledast (\tilde{p}^{(q)}_{-l})^* |_{l = \mu m} $, in \eqref{distMF} were ideal raised cosine impulse response then $\psi^{(\{a_n\},m)}_q$ would depend only on the single symbol $a_m$,  $\psi^{(\{a_n\},m)}_q = \psi^{(a_m)}_q$. However, due to nonlinear distortion, there is a mismatch in matched filtering, which creates ISI over sequence. Nevertheless, it can be observed that $\bar{p}_{l}$ is a decaying function, hence, it is assumed that nonlinearity depends on the sequence of a reduced dimension, ${\bf a}_m = [a_{m-L_q +1},...,a_m,...,a_{m+L_q-1}]^T$. Consequently, received signal at symbol rate is expressed as,

\begin{equation}
     y_m = \varpi_0 a_m +  \sum_{q=1}^{Q-1}   \varpi_q \psi^{({\bf{a}}_m)}_q\left( \sum_{k= -L_q+1}^{ L_q-1 } {{a}}_{m-k}  \bar{p}_{k}^{(q)}  \right). \label{nonlinearISI}
\end{equation}
It can be inferred from \eqref{nonlinearISI} that even for systems subjected the memoryless nonlinearity, distortion signal is a nonlinear function of transmitted sequence. Therefore, sequence detection should be performed in order to decode the signal \cite{Colavolpe}. Consequently, by limiting the memory of the sequence, efficient algorithms can be employed to detect the transmitted sequence. In literature, several methods employing memoryless detection schemes were proposed \cite{ziya,SalehDet}, which can be considered as the special cases of \eqref{nonlinearISI}, where ${\bf a}_m = a_m$,

\begin{equation}
     y_m =  \varpi_0 a_m +  \sum_{q=1}^{Q-1}   \varpi_q \psi^{(a_m)}_q\left(  \bar{p}_{0}^{(q)} \, a_m  \right). \label{eq_isi}
\end{equation}
Eventually, received signal becomes a symbol dependent function and it can be represented by a symbol dependent coefficient, $\varpi^{(a_m)}$ as, $y_m = \varpi^{(a_m)} a_m$, where
\begin{equation}
    \varpi^{(a_m)} = \varpi_0 + \left( \sum_{q=1}^{Q-1} \varpi_q \psi^{(a_m)}_q\left(  \bar{p}_{0}^{(q)} \, a_m  \right)\right) \frac{1}{a_m}.
\end{equation}




%
%

\bibliography{references_bib} 
\bibliographystyle{IEEEtran}

%




\end{document}

%% file: system_block_diagram.tex
\usetikzlibrary{shapes,arrows}
\usetikzlibrary{positioning}
\usetikzlibrary{calc}

\tikzset{%
  block/.style    = {draw, thick, rectangle, minimum height = 1cm,
    minimum width = 3em},
 block1/.style    = {draw, thick, rectangle, minimum height = 6cm,
    minimum width = 3em},
 block2/.style    = {draw, dashed, rectangle, minimum height = 8cm,
    minimum width = 2cm},
  sum/.style      = {draw, circle, node distance = 2cm}, 
  sumq/.style      = {draw, circle, node distance = 2cm}, 
  gain/.style      = {draw, circle, node distance = 2cm}, 
  input/.style    = {coordinate}, 
  output/.style   = {coordinate} 
}

\begin{figure}[ht]\setlength{\unitlength}{0.1in}
\centering
\newcommand{\suma}{$\Sigma$}
\newcommand{\sumaq}{\Large$\Sigma$}
\newcommand{\inte}{$\displaystyle \int$}
\newcommand{\derv}{\huge$\frac{d}{dt}$}
\begin{tikzpicture}[auto, node distance=2cm, >=triangle 45]

\draw

	node [input, name=bn] at (-0,0) {}

	node [align=center, block,font=\tiny\linespread{1.6}\selectfont, right = 0.3in of bn](Mod) { \small QAM \small\\ \small Modulator}

	node [align=center, block, right = 0.3in of Mod](pt) {$p_n$}

	node [input, name=pa0,right = 0.3in of pt ]  {} 
	node [input, name=pa1,above = 0.25in of pa0 ]  {}
	node [input, name=pa2,below = 0.25in of pa0 ]  {}
	node [input, name=pa3,right = 0.35in of pa0 ]  {}

	node [input, name=ant0,right = 0.3in of pa3 ]  {}

	node [input, name=ant1,above = 0.2in of ant0 ]  {}

	node [input, name=ant2,above = 0.15in of ant1 ]  {}
	node [input, name=ant3,right = 0.1in of ant2 ]  {}
	node [input, name=ant4,left = 0.1in of ant2 ]  {}

	node [input, name=xt,right = 0.6in of ant0 ]  {}
	node [align=center, block, right = 0.3in of xt](ch) {$h_n$}
	node [input, name=ch1,right = 0.6in of ch ]  {}
	node [align=center,sumq,below  = 0.5in of Mod] (awgn)   {\suma}
	node [input, name=sumin,left = 0.2in of awgn ]  {}
	node [input, name=sumin1,above = 0.5in of sumin ]  {}
	node [input, name=nt,above = 0.2in of awgn ]  {}
	node [align=center,font=\tiny\linespread{1.6}\selectfont, block, right = 0.3in of awgn](mf) {\small MF \\ \small $p^*_{-n}$}
	node [input, name=yt,right = 0.2in of mf ]  {}
	node [input, name=yts, right = 0.1 in of yt ]  {}
	node [input, name=yts1, above = 0.1 in of yts ]  {}
	node [input, name=yts2, below = 0.2 in of yts ]  {}
	node [input, name=yn, right = 0.05 in of yts ]  {}
	node [input, name=yn1, right = 0.2 in of yn ]  {}
	node [align=center, block, right = 0.3in of yn](DFT) {\small DFT}
	node [align=center, block, right = 0.1in of DFT](FDE) {\small FDE}
	node [align=center, block,font=\tiny\linespread{1.6}\selectfont, below = 0.3in of FDE](chtrain) { \small CSI \\ \small acquisition}
	node [align=center, block, right = 0.1in of FDE](IFFT) {\small IDFT}
	node [align=center, block,font=\tiny\linespread{1.6}\selectfont, right = 0.3in of IFFT](post) {\small NL \small Post-\\ \small Distorter}
	node [align=center, block,font=\tiny\linespread{1}\selectfont, below = 0.3in of post](train) {\small NL \small param.\\ \small Learning}
	node [align=center, block,font=\tiny\linespread{1.6}\selectfont, right = 0.3in of post](demod) {\small Symbol \\ \small Detector.}
	node [input, name=out,right = 0.3in of demod ]  {};

\node[draw=none,fill=none,right = -0.05 in of pa0] {$PA$};	
\draw [->] (bn)-- node { $b_n$} (Mod);
\draw [->] (Mod)-- node { $a_n$} (pt);
\draw [->] (pt)-- node {$x_n$ } (pa0);
\draw [-] (pa0)-- node { } (pa1);
\draw [-] (pa0)-- node { } (pa2);
\draw [-] (pa1)-- node { } (pa3);
\draw [-] (pa2)-- node { } (pa3);

\draw [-] (pa3)-- node { $\tilde{x}_n$} (ant0);
\draw [-] (ant0)-- node { } (ant1);
\draw [-] (ant1)-- node { } (ant3);
\draw [-] (ant1)-- node { } (ant4);
\draw [-] (ant3)-- node { } (ant4);

\draw [->] (xt)-- node { $\tilde{x}_n$} (ch);
\draw [-] (ch)-- node { } (ch1);
\draw [-] (ch1) |- node { } (sumin1);
\draw [-] (sumin1) -- node { } (sumin);
\draw [->] (sumin) -- node { } (awgn);
\draw [->] (nt)-- node { $\nu_n$} (awgn);
\draw [->] (awgn)-- node {} (mf);

\draw [-] (mf)-- node {} (yt);
\draw [-] (yt)-- node {} (yts1);
\draw [->] (yts2)-- node {} (yts);
\node[draw=none,fill=none,below = 0.05 in of yts2] {$n \mu$};	

\draw [->] (yn)-- node {$y_n$} (DFT);
\draw [->] (DFT)-- node {} (FDE);
\draw [->] (FDE)-- node {} (IFFT);
\draw [->] (IFFT)-- node {$z_n$} (post);
\draw [->] (post)-- node {$\tilde {a}_n$} (demod);
\draw [->] (demod)-- node {$\hat {a}_n$} (out);

\draw [loosely dashed,->] (yn1) |- node { } (chtrain);

\draw [->] (chtrain)-- node {$\hat {h}_n$} (FDE);
\draw [->] (train)-- node { } (post);
\end{tikzpicture}
\caption{Transciever structure for SC-FDE transmission system.}
\label{fig:system_block}
\end{figure}

%% file: frame_st.tex
\usetikzlibrary{shapes,arrows}
\usetikzlibrary{positioning}
\usetikzlibrary{calc}

\tikzset{%
  block/.style    = {draw, rectangle, minimum height = 1.cm,
    minimum width =2cm},
 block1/.style    = {draw, rectangle, minimum height = 1.cm,
    minimum width = 1.2cm},
 block2/.style    = {draw, rectangle, minimum height = 1.cm,
    minimum width = 2cm},
 block3/.style    = {draw, rectangle, minimum height = 1.cm,
    minimum width = 0.2cm},
  sum/.style      = {draw, circle, node distance = 2cm}, 
  sumq/.style      = {draw, circle, node distance = 2cm}, 
  gain/.style      = {draw, circle, node distance = 2cm}, 
  input/.style    = {coordinate}, 
  output/.style   = {coordinate} 
}

\begin{figure}[ht]\setlength{\unitlength}{0.12in}
\centering
\newcommand{\suma}{\Large$\Sigma$}
\newcommand{\sumaq}{\Large$\Sigma$}
\newcommand{\inte}{$\displaystyle \int$}
\newcommand{\derv}{\huge$\frac{d}{dt}$}
\begin{tikzpicture}[auto, node distance=2cm, >=triangle 45]

\draw

node [input, name=r_t] at (-0,0) {} 

node [align=center, font=\tiny\linespread{1.6}\selectfont, block, right = 0.3in of r_t](ST) {\footnotesize ST \\ \footnotesize  Training}
node [input, name=stm ,below = 0.25in of ST] {}
node [input, name=stl ,left = 1 cm of stm] {}
node [input, name=str ,right = 1 cm of stm] {}
node [align=center, font=\tiny\linespread{1.6}\selectfont, block1, right = 0in of ST](FT1) {\footnotesize FT \\ \footnotesize  Training}
node [input, name=ftm ,below = 0.25in of FT1] {}
node [input, name=ftl ,left = 0.7 cm of ftm] {}
node [input, name=ftr ,right = 0.7 cm of ftm] {}
node [align=center, block3, right = 0in of FT1](CP1) { \footnotesize CP}
node [align=center, block2, right = 0in of CP1](D1) {\footnotesize Data Stream}
node [input, name=dtm ,below = 0.25in of D1] {}
node [input, name=dtl ,left = 1 cm of dtm] {}
node [input, name=dtr ,right = 1 cm of dtm] {}
node [align=center, block3, right = 0in of D1](CS1) {\footnotesize CS}
node [align=center,font=\tiny\linespread{1.6}\selectfont, block1, right = 0in of CS1](FT1) {\footnotesize FT \\ \footnotesize  Training}
node [align=center, block3, right = 0in of FT1](CP1) { \footnotesize CP}
node [align=center, block2, right = 0in of CP1](D1) {\footnotesize Data  Stream}
node [align=center, block3, right = 0in of D1](CS1) {\footnotesize CS}

node [input, name=outs1 ,right = 0.1in of CS1] {} 
node [input, name=outs2 ,right = 0.75in of CS1] {} ;

\draw [dashed,-] (outs1) -- node { } (outs2);
\draw [<->] (stl) -- node {\footnotesize $N_S$ } (str);
\draw [<->] (ftl) -- node {\footnotesize $N_F$ } (ftr);
\draw [<->] (dtl) -- node {\footnotesize $N_D$ } (dtr);

\end{tikzpicture}
\caption{ Frame structure for the transmition scheme.}
\label{framest}
\end{figure}

%% file: block_diagram.tex
\usetikzlibrary{shapes,arrows}
\usetikzlibrary{positioning}
\usetikzlibrary{calc}

\tikzset{%
  block/.style    = {draw, thick, rectangle, minimum height = 1cm,
    minimum width = 3em},
 block1/.style    = {draw, thick, rectangle, minimum height = 6cm,
    minimum width = 3em},
 block2/.style    = {draw, dashed, rectangle, minimum height = 8cm,
    minimum width = 2cm},
  sum/.style      = {draw, circle, node distance = 2cm}, 
  sumq/.style      = {draw, circle, node distance = 2cm}, 
  gain/.style      = {draw, circle, node distance = 2cm}, 
  input/.style    = {coordinate}, 
  output/.style   = {coordinate} 
}

\begin{figure}[ht]\setlength{\unitlength}{0.12in}
\centering
\newcommand{\suma}{\Large$\Sigma$}
\newcommand{\sumaq}{\Large$\Sigma$}
\newcommand{\inte}{$\displaystyle \int$}
\newcommand{\derv}{\huge$\frac{d}{dt}$}
\begin{tikzpicture}[auto, node distance=2cm, >=triangle 45,scale=0.05]

\draw

	node [input, name=r_t] at (-0,0) {} 

	node [align=center, block, right = 0.3in of r_t](MF) {\small $p^*(-t)$}

	node [ input, name = MFRight, right = 0.3in of MF]{}

	node [input, name=y_t1,above =0.6in of MFRight] {}
	node [input, name=y_t12,right =0.4in of y_t1] {}
	node [input, name=yn1,right =0.15in of y_t12] {}
	node [align=center, block, right = 0.25in of yn1](DFT1) { \small DFT}
	node [input, name=y1sw, above right = 0.1 in of y_t12]{}
	node [input, name=y1ar1,  right = 0.05 in of y_t12]{}
	node [input, name=y1ar2,  below = 0.2 in of y1ar1]{}
	node [align=center, block, right = 0.3in of DFT1](FDE1) { \small FDE}
	node [align=center, block, right = 0.2in of FDE1](IDFT1) { \small IDFT}
	node [align=center, block, right = 0.3in of IDFT1](NR1) { \small NP}
	node [input, name=fdea1,  above = 0.2in of FDE1]{}
	node [input, name=blrn,  right = 0.035in of NR1]{}

	node [input, name=y_t2,above =-0.3in of MFRight] {}
	node [input, name=y_t22,right =0.4in of y_t2] {}
	node [input, name=yn2,right =0.15in of y_t22] {}
	node [align=center, block, right = 0.25in of yn2](DFT2) { \small DFT}
	node [input, name=y2sw, above right = 0.1 in of y_t22]{}
	node [input, name=y2ar1,  right = 0.05 in of y_t22]{}
	node [input, name=y2ar2,  below = 0.2 in of y2ar1]{}
	node [align=center, block, right = 0.3in of DFT2](FDE2) { \small FDE}
	node [align=center, block, right = 0.2in of FDE2](IDFT2) { \small IDFT}
	node [align=center, block, right = 0.3in of IDFT2](NR2) { \small NP}
	node [input, name=fdea2,  above = 0.2 in of FDE2]{}
	node [input, name=blrn2,  right = 0.08in of NR2]{}

	node [input, name=y_tn,above = -1.4in of MFRight] {}
	node [input, name=y_tn2,right =0.4in of y_tn] {}
	node [input, name=ynn,right =0.15in of y_tn2] {}
	node [align=center, block, right = 0.25in of ynn](DFTn) { \small DFT}
	node [input, name=ynsw, above right = 0.1 in of y_tn2]{}
	node [input, name=ynar1,  right = 0.05 in of y_tn2]{}
	node [input, name=ynar2,  below = 0.2 in of ynar1]{}
	node [align=center, block, right = 0.3in of DFTn](FDEn) { \small FDE}
	node [align=center, block, right = 0.2in of FDEn](IDFTn) { \small IDFT}
	node [align=center, block, right = 0.3in of IDFTn](NRn) { \small NP}
	node [input, name=fdean,  above = 0.2 in of FDEn]{}
		node [input, name=blrn3,  right = 0.3in of NRn]{}

	node [align=center, block, below = 2in of MF](corr) { \small Correlator}
	node [input, name=corr1,  below = 0.5 in of ynar1]{}

	node [input, name=yd1 ,right =0.46in of NR1] {}
	node [input, name=yd2 ,right =0.46in of NR2] {}
	node [input, name=ydn ,right =0.46in of NRn] {}

	node [align=center, block2, below = -1.8in of FDE2](FDEALL) {}
	node [align=center,font=\tiny\linespread{1.6}\selectfont, block, below = 0.3in of FDEALL](CSI) { \small Symbol rate\\ \small CSI acqusition}
	
	node [input, name=yiin ,left =0.75 of CSI] {}
	
	node [input, name=nrt ,below = 0.76in of NRn] {}
	node [align=center, block,font=\tiny\linespread{1.6}\selectfont, left = 0.2 in of nrt](paramLe) { \small NP \\ \small Training}
	node [input, name=ST1,  below = 0.2 in of paramLe]{}
    
	node [input, name=ST2,  above = 0.2 in of NR1]{}
    node [input, name=ST4,  below = 0.2 in of CSI]{}

	node [align=center, block1, below right = -3 and 1.2 of NR2](Detect) { \small Distortion \\\small Aware \\ \small Bussgang- \\ \small aided  \\ \small detector}
	node [align=center,font=\tiny\linespread{1.6}\selectfont, block, below =0.2in of Detect](BussLearn) {\small Detector \\\small Parameter \\\small Learning}
	node [input, name=ST3,  below = 0.2 in of BussLearn]{}

	node [input, name=est,  right = 0.3 in of Detect]{};

\node[draw=none,fill=none,below = 0.22 in of y1ar1] {\small $t=nT$}; 
\node[draw=none,fill=none,below = 0.22 in of y2ar1] {\small $t=nT+\frac{T}{\mu}$}; 
\node[draw=none,fill=none,below = 0.22 in of ynar1] {\small $t=nT+\frac{T(\mu-1)}{\mu}$}; 

\node[draw=none,fill=none,above = 0.16 in of FDE1] {${\bf {\hat{h}}}^{(1)}$}; 
\node[draw=none,fill=none,above = 0.16 in of FDE2] {${\bf {\hat{h}}}^{(2)}$}; 
\node[draw=none,fill=none,above = 0.16 in of FDEn] {${\bf {\hat{h}}}^{(\mu)}$}; 

\node[draw=none,fill=none,below = 0.16 in of paramLe] {\small ST Training}; 

\node[draw=none,fill=none,below = 0.16 in of CSI] {\small FT Training};

\node[draw=none,fill=none,below = 0.16 in of BussLearn] {\small FT Training}; 

\draw [->] (r_t) -- node {$r(t)$} (MF) ;
\draw [-] (MF) -- node { $y(t)$} (MFRight) ;
\draw [-] (MFRight) -| node { } ( y_t1) ;
\draw [-] (y_t1) -- node { } ( y_t12) ;
\draw [-] (y_t12) -- node { } ( y1sw) ;
\draw [->] (y1ar2) -- node {} (y1ar1) ;
\draw [->] (yn1) -- node {\small $y_n^{(1)}$} (DFT1);
\draw [->] (DFT1) -- node {} (FDE1);
\draw [->] (FDE1) -- node {} (IDFT1);
\draw [->] (IDFT1) -- node {$z_n^{(1)}$} (NR1);
\draw [->] (NR1) -- node { $\tilde{a}_n^{(1)}$} (yd1) ;
\draw [->] (fdea1) -- node {} (FDE1);

\draw [-] (y_t2) -- node { } ( y_t22) ;
\draw [-] (y_t22) -- node { } ( y2sw) ;
\draw [->] (y2ar2) -- node {} (y2ar1) ;
\draw [->] (yn2) -- node {\small $y_n^{(2)}$} (DFT2);
\draw [->] (DFT2) -- node {} (FDE2);
\draw [->] (FDE2) -- node {} (IDFT2);
\draw [->] (IDFT2) -- node {$z_n^{(2)}$} (NR2);
\draw [->] (NR2) -- node {$\tilde{a}_n^{(2)}$ } (yd2);
\draw [->] (fdea2) -- node {} (FDE2);

\draw [-] (MFRight) -| node { } ( y_tn) ;
\draw [-] (y_tn) -- node { } ( y_tn2) ;
\draw [-] (y_tn2) -- node { } ( ynsw) ;
\draw [->] (ynar2) -- node {} (ynar1) ;
\draw [->] (ynn) -- node {\small $y_n^{(\mu)}$} (DFTn);
\draw [->] (DFTn) -- node {} (FDEn);
\draw [->] (FDEn) -- node {} (IDFTn);
\draw [->] (IDFTn) -- node {$z_n^{(\mu)}$} (NRn);
\draw [->] (NRn) -- node {$\tilde{a}_n^{(\mu)}$ } (ydn) ;
\draw [->] (fdean) -- node {} (FDEn);

\draw [->] (CSI) -- node {${\bf \hat h}^{(i)}$} (FDEALL);

\draw [->] (ST1) -- node { } (paramLe);
\draw [->] (ST3) -- node { } (BussLearn);
\draw [->] (ST4) -- node { } (CSI);

\draw [dashed,-] (paramLe) -| node { } (NRn);
\draw [dashed,-] (NRn) -- node { } (NR2);
\draw [dashed,-] (NR2) -- node { } (NR1);
\draw [dashed,->] (NR1) -- node { } (ST2);

\draw [->] (BussLearn) -- node {${ \boldsymbol \beta }, \; {\bf R}_{\eta}$} (Detect);
\draw [->] (Detect) -- node { $\hat{a}_n$} (est);

\draw [->] (corr) -| node { } (corr1);
\draw [loosely dashed,->] (blrn) |- node { } (BussLearn);
\draw [loosely dashed,->] (blrn2) |- node { } (BussLearn);
\draw [loosely dashed,->] (blrn3) |- node { } (BussLearn);

\draw [->] (yiin) -- node {$y_n^{(i)}$} (CSI);

\end{tikzpicture}
\caption{Block diagram for proposed bank of FDE's based detector.}
\label{bankFde}
\end{figure}

%% file: NNArch.tex
\usetikzlibrary{shapes,arrows}
\usetikzlibrary{positioning}
\usetikzlibrary{calc}

\tikzset{%
  block/.style    = {draw, thick, rectangle, minimum height = 0.75cm,
    minimum width = 0.75cm},
delayb/.style    = {draw, thick, rectangle, minimum height = 2cm,
    minimum width = 3em},
 block2/.style    = {draw, dashed, rectangle, minimum height = 8cm,
    minimum width = 2cm},
  sum/.style      = {draw, circle, node distance = 2cm}, 
  sumq/.style      = {draw, circle, node distance = 1cm}, 
  gain/.style      = {draw, circle, node distance = 2cm}, 
  input/.style    = {coordinate}, 
  output/.style   = {coordinate} 
}

\begin{figure}[ht]\setlength{\unitlength}{0.12in}
\centering
\newcommand{\delaya}{\Large$z^{-1}$}
\newcommand{\sumaq}{\small$\Sigma$}
\newcommand{\inte}{$\displaystyle \int$}
\newcommand{\derv}{\huge$\frac{d}{dt}$}
\begin{tikzpicture}[auto, node distance=5cm, >=triangle 60]

\draw

	node [input, name=r_n] at (-0,0) {} 

	node [input, name=r_n1, right = 0.5in of r_n]{}
	node [input, name=r_n2, right = 0.375in of r_n1]{}
	node [align=center, block, below = 0.2in of r_n1](delay1) {$z^{-1}$}
	node [align=center, block, below = 0.3in of delay1](delay3) {$z^{-1}$}
	
	node [input, name=d1n, right = 0.2in of delay1]{}
	node [input, name=d3n, right = 0.2in of delay3]{}
	
	node [input, name=i_n, below =1.5in of r_n]{}

	node [input, name=i_n1, right = 0.5in of i_n]{}
	node [input, name=i_n2, right = 0.375in of i_n1]{}
	node [align=center, block, below = 0.2in of i_n1](delay4) {$z^{-1}$}
	node [align=center, block, below = 0.3in of delay4](delay6) {$z^{-1}$}

	node [input, name=d4n, right = 0.2in of delay4]{}
	node [input, name=d6n, right = 0.2in of delay6]{}

	node [align=center,sumq,right = 0.6in of delay1] (layera)   {\sumaq}
	node [align=center,sumq,right = 0.6in of delay3] (layerc)   {\sumaq}
	node [align=center,sumq,below = 1.in of layerc] (layerb)   {\sumaq}

	node [align=center, block, right = 0.3in of layera](acta) {$g(\cdot)$}
	node [align=center, block, right = 0.3in of layerb](actb) {$g(\cdot)$}
	node [align=center, block, right = 0.3in of layerc](actc) {$g(\cdot)$}
	
	node [input, name=ca, right = 0.2in of acta]{}
	node [input, name=cb, right = 0.2in of actb]{}
	node [input, name=cc, right = 0.2in of actc]{}

	node [align=center,sumq,below right = 0.2in of ca] (outa)   {\sumaq}
	node [align=center,sumq,above right = 0.2in of cb] (outb)   {\sumaq}

	node [input, name=inphase, right = 0.6in of outa]{}
	node [input, name=quadrature, right = 0.6in of outb]{};

\draw [-] (r_n) -- node {$\operatorname{Re}\{ z_n\}$} (r_n1) ;
\draw [->] (r_n1) -- node { } (delay1) ;
\draw [dashed,->] (delay1) -- node { } (delay3) ;

\draw [-] ( r_n1) -- node { } ( r_n2) ;
\draw [-] (delay3) -- node { } (d3n) ;

\draw [-] (i_n) -- node {$\operatorname{Im}\{ z_n\}$} (i_n1) ;
\draw [->] (i_n1) -- node { } (delay4) ;
\draw [dashed,->] (delay4) -- node { } (delay6) ;

\draw [-] ( i_n1) -- node { } ( i_n2) ;
\draw [-] (delay6) -- node { } (d6n) ;

\node [input, name=weighta, circle, right = 0.1 in of r_n2]{\small $w_{1,1}$} ;
\node [input, name=weighta, circle, right = 0.2 in of d6n]{ \small$w_{L_1,4M-2}$} ;

\draw [->] (r_n2) -- node { } (layera) ;
\draw [->] (i_n2) -- node { } (layera) ;
\draw [->] (d3n) -- node { } (layera) ;
\draw [->] (d6n) -- node { } (layera) ;

\draw [->] (r_n2) -- node { } (layerc) ;
\draw [->] (i_n2) -- node { } (layerc) ;
\draw [->] (d3n) -- node { } (layerc) ;
\draw [->] (d6n) -- node { } (layerc) ;

\node [input, name=biasaa, circle,radius=0.05,above right = 0.2in of layera]{ $b_{1,1}$} ;
\node [input, name=biasab, circle,radius=0.05,above right = 0.2in of layerb]{ $b_{1,L_1}$} ;
\node [input, name=biasae, circle,radius=0.05,above right = 0.2in of layerc]{ $b_{1,2}$} ;

\node [input, name=biasac, circle,radius=0.05,above right = 0.2in of outa]{ $b_{2,I}$} ;
\node [input, name=biasad, circle,radius=0.05,above right = 0.2in of outb]{ $b_{2,Q}$} ;

\draw [-] (biasaa) -- node { } (layera) ;
\draw [-] (biasab) -- node { } (layerb) ;
\draw [-] (biasae) -- node { } (layerc) ;

\draw [-] (biasac) -- node { } (outa) ;
\draw [-] (biasad) -- node { } (outb) ;

\draw [loosely dashed] (layerc) -- node { } (layerb) ;

\draw [->] (r_n2) -- node { } (layerb) ;
\draw [->] (i_n2) -- node { } (layerb) ;
\draw [->] (d3n) -- node { } (layerb) ;
\draw [->] (d6n) -- node { } (layerb) ;

\draw [->] (layera) -- node { } (acta) ;
\draw [->] (layerb) -- node { } (actb) ;
\draw [->] (layerc) -- node { } (actc) ;

\draw [loosely dashed] (actc) -- node { } (actb) ;

\draw [-] (acta) -- node { } (ca) ;
\draw [-] (actb) -- node { } (cb) ;
\draw [-] (actc) -- node { } (cc) ;

\draw [->] (ca) -- node { } (outa) ;
\draw [->] (ca) -- node { } (outb) ;

\draw [->] (cb) -- node { } (outa) ;
\draw [->] (cb) -- node { } (outb) ;

\draw [->] (cc) -- node { } (outa) ;
\draw [->] (cc) -- node { } (outb) ;

\draw [->] (outa) -- node { $\operatorname{Re}\{ \tilde{a} \}$} (inphase) ;
\draw [->] (outb) -- node {  $\operatorname{Im}\{ \tilde{a} \}$} (quadrature) ;

\end{tikzpicture}
\caption{ ARVTDNN architecture.}
\label{fig:ARVTDNN}
\end{figure}